\documentclass[12pt,english]{article}
\usepackage{amsmath}
\usepackage{graphicx}
\usepackage{setspace}
\usepackage{amssymb}
\usepackage{a4wide}
\usepackage{cite} % [1-5] instead of [1,2,3,4,5]
\usepackage{babel}

\begin{document}
\date{\mbox{ }}

%\title{Weierstrass meets Enriques}
\title{ 
{\normalsize     
HD-THEP-09-16\hfill\mbox{}\\
%November 2008\hfill\mbox{}\\
}
\vspace{1cm}
\bf 
Weierstrass meets Enriques\\[8mm]}

\author{A.~P.~Braun, R.~Ebert, A.~Hebecker, R.~Valandro\\[2mm]
{\normalsize\itshape  Institut f\"ur Theoretische Physik, Universit\"at Heidelberg,}\\
{\normalsize\itshape Philosophenweg 16-19, 69120 Heidelberg, Germany}
\thanks{{\ttfamily a.braun}, {\ttfamily a.hebecker}, {\ttfamily r.valandro} {\ttfamily
    @thphys.uni-heidelberg.de}, {\ttfamily email@rainerebert.de}%
}}
% \author{A. P. Braun, R. Ebert, A. Hebecker and R. Valandro
% \thanks{E-Mail: a.braun@thphys.uni-heidelberg.de,
% email@rainerebert.de, a.hebecker@thphys.uni-heidelberg.de, r.valandro@thphys.uni-heidelberg.de%
% }}

\maketitle

\begin{abstract}
We study in detail the degeneration of $K3$ to $T^{4}/\mathbb{Z}_{2}$. 
We obtain an explicit embedding of the lattice of collapsed cycles of $T^{4}/\mathbb{Z}_{2}$
into the lattice of integral cycles of $K3$ in two different ways. Our first method exploits the
duality to the heterotic string on $T^3$. This allows us to describe the degeneration in terms of Wilson
lines. Our second method is based on the blow-up of $T^{4}/\mathbb{Z}_{2}$. From this blow-up, we directly
construct the full lattice of integral cycles of $K3$. Finally, we use our results to describe the action of the 
Enriques involution on elliptic $K3$ surfaces, finding that a Weierstrass model description is consistent 
with the Enriques involution only in the F-theory limit.
\end{abstract}

\newpage{}

\newpage{}
\tableofcontents

\section{Introduction}

In the study of string-theory compactifications, the geometric understanding 
of the cycle structure of complex manifold plays a central role. Examples are 
F-theory models with fluxes (see e.g. \cite{Denef:2008wq} for a review and 
\cite{dw08,bhv08,Blumenhagen:2008zz,Andreas:2009uf,Marsano:2009ym,Collinucci:2008pf,Blumenhagen:2009gk,Collinucci:2008zs,Blumenhagen:2009up}
for recent work) and blow-ups of heterotic orbifolds (see 
e.g.\cite{Nibbelink:2008tv,Heter1,Heter2,Heter3,Heter4}). One of the simplest relevant geometries, 
which may also play a role as a building block in more complex models, is the $K3$ 
surface~\cite{Aspinwall:1996mn,Aspinwall:1995vk,Aspinwall:1997eh}.

Shrinking sixteen two-spheres in $K3$, the surface 
develops sixteen $A_1$ singularities. This corresponds to the 
$T^{4}/\mathbb{Z}_{2}$ orbifold limit. To describe this degeneration in 
detail, we need to know which two-spheres shrink. The answer to this question
represents our central result: We construct an embedding of the cycles of 
$T^{4}/\mathbb{Z}_{2}$, including the lattice $A_1^{\oplus 16}$ of collapsed cycles, 
into the lattice $\Gamma_{3,19}=U^{\oplus 3}\oplus (-E_8)^{\oplus 2}$ of integral cycles 
of $K3$. This embedding yields an elegant description of $\Gamma_{3,19}$ making all the 
symmetries of $T^{4}/\mathbb{Z}_{2}$ manifest. It also allows for a rather intuitive understanding 
of the cycle structure and certain regions of the moduli space of $K3$, which 
is based on the possibility to visualize $T^{4}/\mathbb{Z}_{2}$ using a 
four-dimensional hypercube. 

Furthermore, we are interested in the action of the Enriques involution\footnote{The Enriques 
involution is a fixed-point free holomorphic involution of $K3$ which is non-symplectic, i.e. it 
projects out the holomorphic two-form \cite{peters,key-67}. It yields the Enriques surface as the 
quotient space.} on elliptic $K3$ surfaces, especially its compatibility with the description of $K3$ by
a Weierstrass model. The Weierstrass model is of particular interest as it is commonly used in 
the context of F-theory \cite{Vafa:1996xn,Sen:1997bp}. It is known that $T^{4}/\mathbb{Z}_{2}$ allows an Enriques involution \cite{key-65}. 
Using the results of the first part of this paper, we show how to deform $T^{4}/\mathbb{Z}_{2}$ to a $K3$ given 
by the standard Weierstrass form. It turns out that this deformation is not consistent with the {\it holomorphicity}
of the Enriques involution, the obstacle being the single distinguished section of the Weierstrass model\footnote{
A freely acting $\mathbb{Z}_2$-symmetry of the real metric manifold still exists, but it is not holomorphic
in the complex structure of the Weierstrass model.}. This problem does not arise for elliptic $K3$
surfaces that are given in non-standard Weierstrass form, e.g. one with two distinguished 
sections \cite{Klemm:1996ts,Berglund:1998va}. In the F-theory limit, however, also the usual Weierstrass form 
becomes symmetric under the Enriques involution.

This paper is organized as follows:

In Sects. \ref{sec2} and \ref{sec3}, we explicitly work out the 
equivalence between the resolution of singularities of $K3$ and Wilson 
line breaking of $E_8\times E_8$. In particular, we show that the relevant 
breaking of $E_8$ to $SU(2)^8$ is highly symmetric: It is achieved by 
three Wilson lines which are all equivalent through automorphisms of the 
$E_8$ lattice.

Sect.~\ref{sec4} combines the results of the previous two sections and
identifies the integral cycles of $K3$ which shrink to produce the sixteen
$A_1$ singularities. Furthermore, we reproduce the known action of the 
Enriques involution on $T^{4}/\mathbb{Z}_{2}$ \cite{key-65} from its action on 
$H_{2}(K3,\mathbb{Z})$ as given in the mathematics literature \cite{peters}.

In Sect.~\ref{sec5}, we describe $K3$ and in particular
$T^{4}/\mathbb{Z}_{2}$ as a double cover of $\mathbb{P}^1\times \mathbb{P}^1$. 
This description nicely displays holomorphic sections and shows which 
of the singularities they hit.

We then construct the full lattice $H_{2}(K3,\mathbb{Z})$ in a 
blow-up of $T^{4}/\mathbb{Z}_{2}$ in Sect.~\ref{sec6}. Our starting point are 
the six even cycles of $T^{4}$ and the sixteen exceptional divisors emerging 
in the blow-up of the singularities. While these cycles span 
$H_{2}(K3,\mathbb{R})$ as a real vector space, they do not form an integral 
basis of $H_{2}(K3,\mathbb{Z})$. We construct the extra integral cycles which 
complete the lattice $U(2)^{\oplus3} \oplus A_{1}^{\oplus16}$ to 
$U^{\oplus3}\oplus(-E_{8})^{\oplus2}$. The structure of this complete lattice
can be nicely displayed in terms of a four-dimensional cube. This provides
an intuitive geometrical picture of the cycles of $K3$.

We demonstrate the equivalence between the two embeddings of $A_{1}^{\oplus16}$
into $\Gamma_{3,19}$ in Sect.~\ref{sec7} by finding an explicit map between 
them.

In Sect.~\ref{sec8} we relate the action of the Enriques involution on
$T^{4}/\mathbb{Z}_{2}$ to its action on the lattice of integral cycles of 
$K3$. We proceed by showing that elliptic $K3$ surfaces described by the standard
Weierstrass model do not allow an Enriques involution. It turns out that 
the Enriques involution requires the existence of at least two
holomorphic sections (which are mapped to each other).

In Sect.~\ref{sec9} we finally discuss the F-theory limit of
$T^{4}/\mathbb{Z}_{2}$ and of elliptic $K3$s described by a Weierstrass model with one or
two distinguished sections. Even though these spaces are different and correspond to different 
M-theory compactifications, they yield equivalent models in the F-theory 
limit in which the fibre of the elliptic fibrations is collapsed\footnote{This is clear since our models have the same constant $\tau$
as a function of the base, and this fact is the only feature of the fibration that is relevant in F-theory, see e.g. 
\cite{de Boer:2001px}}. This means in particular that in this limit the standard Weierstrass model becomes symmetric under the Enriques involution.

%\newpage

\section{The lattice \boldmath$E_{8}$ and its sublattice \boldmath$A_{1}^{\oplus8}$}\label{sec2}

The $E_{8}$ root lattice is the unique even unimodular lattice of rank 8.
Any element takes the form $\alpha=q_{I}E_{I}$, where $\{E_{I}\}_{I=1,...,8}$
is a basis of $\mathbb{R}^{8}$ satisfying $E_{I}\cdot E_{J}=-\delta_{IJ}$.
The coordinates have to be all integer or half-integer and must 
fulfill $\sum_{I=1,...,8}q_{I}=2\mathbb{Z}$ \cite{key-49}. Thus the lattice is 
generated by vectors of the type
\[ (\pm1,\pm1,0,0,0,0,0,0) \qquad \mbox{and} \qquad (\pm\frac{1}{2},\pm\frac{1}{2},\pm\frac{1}{2},\pm\frac{1}{2},\pm\frac{1}{2},\pm\frac{1}{2},\pm\frac{1}{2},\pm\frac{1}{2}),\]
where the second type of vectors must have an even number of minus signs.
We choose the (non-unique) set of 8 simple roots
\begin{align}
\alpha_{1} & =\frac{1}{2}E_{1}+\frac{1}{2}E_{2}+...+\frac{1}{2}E_{8} & \alpha_{5} & =-E_{4}+E_{5}\nonumber \\
\alpha_{2} & =-E_{7}-E_{8} & \alpha_{6} & =-E_{3}+E_{4}\nonumber \\
\alpha_{3} & =-E_{6}+E_{7} & \alpha_{7} & =-E_{2}+E_{3}\nonumber \\
\alpha_{4} & =-E_{5}+E_{6} & \alpha_{8} & =-E_{7}+E_{8}.\label{eq:e8rootsystem}\end{align}

\begin{figure}[tt]
\begin{centering}
\includegraphics[width=7cm]{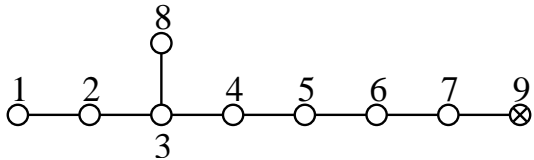}
\par\end{centering}

\caption{The extended Dynkin diagram of $E_{8}$.}\label{fig1}

\end{figure}

The structure of this basis is encoded in the \textit{Dynkin diagram}
of $E_{8}$. The \textit{extended Dynkin diagram}
is obtained by adding the (linearly dependent and thus non-simple) highest root \cite{helgason} (see Fig.~\ref{fig1}).
\begin{align}
\alpha_{9} & =-2\alpha_{1}-4\alpha_{2}-6\alpha_{3}-5\alpha_{4}-4\alpha_{5}-3\alpha_{6}-2\alpha_{7}-3\alpha_{8}
  =-E_{1}+E_{2}.\label{eq:e8highestroot}\end{align}
The coefficients in this expansion are known as the \textit{Coxeter labels}.

The reflections in the hyperplanes orthogonal to the 240 roots are
symmetries of the $E_{8}$ root lattice and generate the Weyl group
of type $E_{8}$. Its order is given by $4!\cdot6!\cdot8!=696729600$~\cite{key-49}. 
The $E_8$ Weyl group contains a subgroup of order $8!\cdot 2^7$ consisting of all permutations of the coordinates and all even sign changes. This subgroup is the Weyl group of type $D_8$. The full $E_8$ Weyl group is generated by this subgroup and the block diagonal matrix ${\cal H}_4\oplus {\cal H}_4$ where ${\cal H}_4$ is the Hadamard matrix
\begin{equation}
    {\cal H}_4 = \tfrac{1}{2}\left(\begin{smallmatrix} 1 & 1 & 1 & 1\\ 1 & -1 & 1 & -1\\ 1 & 1 & -1 & -1\\ 1 & -1 & -1 & 1\\ \end{smallmatrix}\right)\:.
\end{equation}

In gauge field theories based on a certain group, the symmetry can be broken by introducing 
Wilson lines associated with non-contractible loops of the underlying space-time geometry. This is
in one-to-one correspondence with Dynkin's method for finding maximal subgroups by deleting 
nodes in the extended Dynkin diagram.

The action of a Wilson line in $E_8$ (viewed as a vector in $\mathbb{R}^{8}$) on a root $\alpha$ is
\begin{equation}
\alpha\mapsto e^{2\pi i\alpha\cdot W}\alpha\:.\end{equation}
To find the sublattice of $E_{8}$ which corresponds to deleting a simple root $\alpha_{i}$,
we choose a Wilson line $W$ satisfying (see, e.g., \cite{key-51,key-51bis})
\begin{equation}
 \alpha_{i}\cdot W\not\in\mathbb{Z} \qquad\qquad \mbox{and}\qquad\qquad \alpha_{j}\cdot W\in\mathbb{Z} \qquad{\rm for }\: j\in\{1,...,9\}\setminus
\lbrace i \rbrace\:.
\end{equation}
Requiring this transformation to be a symmetry of the root
lattice, we are left with the sublattice of roots satisfying 
$\alpha\cdot W\in\mathbb{Z}$ \cite{key-55,key-54}.
In the following we will show that
\begin{equation}
W^{1}=(1,0^7), \hspace{.5cm}W^{2}=(0^4,-\frac{1}{2}^4)\hspace{.5cm}\mbox{and}\hspace{.5cm}
W^{3}=(0^2,-\frac{1}{2},\frac{1}{2},0^2,-\frac{1}{2},\frac{1}{2})\label{eq5}
\end{equation}
take us from $E_{8}$ to $A_{1}^{\oplus8}$.

It is easy to see that these three Wilson lines are equivalent, i.e. they are related by a
Weyl reflection\footnote{For example, if we apply ${\cal H}_4\oplus{\cal H}_4$ and the following element of the Weyl subgroup of type $D_8$, $(E_1,E_2,E_3,E_4,E_5,E_6,E_7,E_8)\mapsto(-E_3,E_8,E_4,-E_7,-E_1,E_6,E_2,-E_5)$, we get $W^1\mapsto W^3$, $W^2\mapsto W^1$ and $W^3\mapsto W^2$.}.

Let us start with $W^{1}$. This Wilson line removes $\alpha_{1}$,
giving us the Dynkin diagram~of~$D_{8}$.
\begin{figure}
\begin{centering}
\includegraphics[width=6cm]{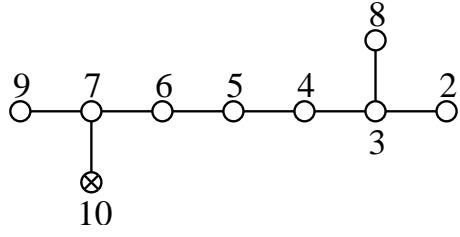}
\par\end{centering}
\caption{The extended Dynkin diagram of $D_{8}$.}\label{ExtDynD8}
\end{figure}
Adding the highest root of the $D_8$ lattice,
\begin{equation}
\alpha_{10}=-\alpha_{2}-2\alpha_{3}-2\alpha_{4}-2\alpha_{5}-2\alpha_{6}-2\alpha_{7}-\alpha_{8}-\alpha_{9}=E_{1}+E_{2}\:.\nonumber
\end{equation}
we obtain the extended Dynkin diagram of $D_8$ (see Fig.~\ref{ExtDynD8}). 

\begin{figure}[tt]
\begin{centering}
\includegraphics[width=2cm]{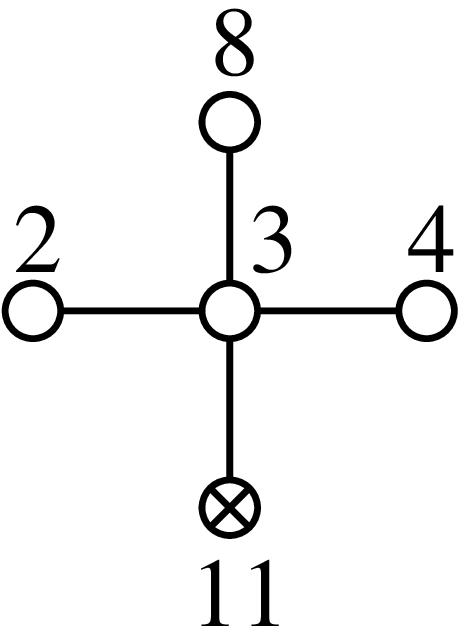}\hspace{2cm}\includegraphics[width=2cm]{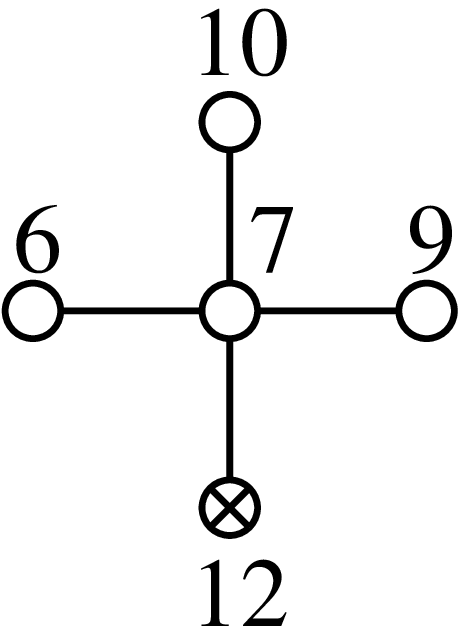}
\par\end{centering}

\caption{Twice the extended Dynkin diagram of $D_{4}$.}\label{fig2D4}

\end{figure}

Next, $W^{2}$ removes the node corresponding to $\alpha_{5}$. We are
left with two copies of the Dynkin diagram of $D_{4}$ (see Fig.~\ref{fig2D4}), which we
extend by their respective highest roots
\begin{equation}
\alpha_{11}=-\alpha_{2}-2\alpha_{3}-\alpha_{4}-\alpha_{8}=E_{5}+E_{6} \qquad \mbox{and} \qquad%\nonumber
%\end{equation}
%and 
%\begin{equation}
\alpha_{12}=-\alpha_{6}-2\alpha_{7}-\alpha_{9}-\alpha_{10}=-E_{3}-E_{4}\:.\nonumber
\end{equation}

Finally, $W^{3}$ removes $\alpha_{3}$
and $\alpha_{7}$, leaving us with 8 unconnected nodes corresponding
to the $A_{1}^{\oplus8}$ sublattice of $E_{8}$.\footnote{Here, the removed nodes are two instead of one; this is because the Wilson line acts on two simple groups.} 
The remaining simple roots are:
\begin{align}
\alpha_{2} &=-E_{7}-E_{8} &\qquad &\alpha_{4} =-E_{5}+E_{6} &\qquad & \alpha_{6} =-E_{3}+E_{4}&\qquad & \alpha_{8}  =-E_{7}+E_{8}\nonumber \\
\alpha_{9} & =-E_{1}+E_{2}&\qquad & \alpha_{10}  =E_{1}+E_{2}&\qquad & \alpha_{11} =E_{5}+E_{6}&\qquad & \alpha_{12} =-E_{3}-E_{4}\end{align}

\section{Moduli space of \boldmath$K3$ and Wilson line breaking}\label{sec3}

Up to diffeomorphisms, $K3$ is the only non-trivial compact Calabi-Yau twofold.\footnote{For a comprehensive review of $K3$, see \cite{Aspinwall:1996mn}.} Its second homology class $H_{2}(K3,\mathbb{Z})$, equipped with the natural metric given by the intersection 
numbers between cycles, is an even, self-dual lattice with signature $(3,19)$, commonly denoted by $\Gamma_{3,19}$. There is a basis
of $H_{2}(K3,\mathbb{Z})$ such that the matrix formed by the inner products of the basis vectors reads
\begin{equation}
U\oplus U\oplus U\oplus(-E_{8})\oplus(-E_{8}),\label{eq:intersectionmatrix}\end{equation}
where $E_{8}$ is the positive definite Cartan matrix of $E_{8}$ and
\[
U=\left(\begin{array}{cc}
0 & 1\\
1 & 0\end{array}\right).\]

We will denote the basis vectors spanning the three $U$ blocks by
$e_{i}$ and $e^{i}$, $i=1,2,3$. Accordingly, $e^{i}\cdot e_{j}=\delta_{j}^{i}$.
Using the notation introduced in the last section for the $E_8$ lattice, any integral 2-cycle can 
now be written as
\begin{equation}
p^{i}e^{i}+p_{i}e_{i}+q_{I}E_{I},\label{intk3}
\end{equation}
where $i=1,2,3$ and $I=1,...,16$. The $p_{i}$ as well as the $p^{i}$
are all integers, while the $q_{I}$ fulfill the relations $\sum_{I=1,...,8}q_{I}=2\mathbb{Z}$
and $\sum_{I=9,...,16}q_{I}=2\mathbb{Z}$ and furthermore have to be
\textit{all} integer or \textit{all} half-integer in each of the two
$E_{8}$ blocks.

A point in the moduli space $M_{K3}$ of $K3$ is chosen by fixing the overall
volume of $K3$ and a positive signature 3-plane $\Sigma$ in
$H_{2}(K3,\mathbb{\mathbb{R}})\cong\mathbb{R}^{3,19}$. 
We choose three real 2-cycles $\omega_{i}\in H_{2}(K3,\mathbb{\mathbb{R}})$,
$i=1,2,3$, which fulfill the constraints $\omega_{i}\cdot\omega_{j}=\delta_{ij}$ and span $\Sigma$. 
A real K\"ahler form $j$ and a holomorphic two-form $\omega$ for the $K3$ surface specified by $\Sigma$ 
are then given by $j=\sqrt{2\cdot \mbox{Vol}(K3)}\cdot\omega_{3}$ and $\omega=\omega_{1}+i\omega_{2}$, 
respectively\footnote{Here and below, we use the same character for a 2-form, its associated
cohomology class and its Poincar\'{e}-dual 2-cycle.}.

The \textit{roots} of $\Gamma_{3,19}$ are defined as the elements of $H_{2}(K3,\mathbb{Z})$ with self-intersection
$-2$.
If a root becomes orthogonal to $\Sigma$, the $K3$ surface develops a singularity since the corresponding 2-cycle shrinks%
\footnote{The volume of a 2-cycle $\gamma$ is proportional to its projection on
$\Sigma$.}.

For example, a $\mathbb{C}^{2}/\mathbb{Z}_{2}$ singularity (also called $A_{1}$
singularity) arises if a single root shrinks. In general, the
singularities which can occur are of A-D-E type and are specified by the
simple roots in the orthogonal complement of $\Sigma$. The intersection matrix of the
cycles corresponding to these roots can be shown to always be minus the Cartan matrix of
some A-D-E group \cite{peters,Aspinwall:1996mn}. This group uniquely determines the A-D-E-type singularity
of the $K3$ surface given by $\Sigma$.

The $E_8\times E_8$ point in moduli space is realized when $\Sigma$ is located in the $U^{\oplus3}$ block spanned by $e_i,e^i$ ($i=1,2,3$)\footnote{Accordingly, $\omega_{i}^{E_{8}\times E_{8}}=a_{i}^{j}e_{j}+b_{k}^{i}e^{k}$, $i,j,k=1,2,3$, for real numbers $a_{i}^{j}$ and $b_{k}^{i}$ s.t. $\omega_{i}\cdot\omega_{j}=\delta_{ij}$.}.
%thus the simple roots of $\Lambda$, the sublattice of $\Gamma_{3,19}$ spanned by the roots in the orthogonal complement of $\Sigma$, give twice the Dynkin diagram of $E_{8}$.
Rotating the plane into the $E_8\times E_8$ block changes the singularity and eventually gives rise to a smooth $K3$. Singularities which may still be present after this rotation correspond to subgroups of $E_8\times E_8$. As we explain in detail in the following, one can relate the symmetry breaking by Wilson lines described in Sect.~\ref{sec2} to the rotation of the $\Sigma$ plane in $H_2(K3,\mathbb{R})$.

We first consider the rotation of $\Sigma$ from a point with $E_8\times E_8$ singularity to a point with $D_8\times E_8$ singularity. The Wilson line that realizes this breaking is $W_{I}=(1,0,0,0,0,0,0,0)$ (see Sect.~\ref{sec2}). This identifies a vector  $W=W_{I}E_{I}=(1,0,0,0,0,0,0,0)$ in the subspace of $H_{2}(K3,\mathbb{R})$ that corresponds to the first $E_{8}$ block in \eqref{eq:intersectionmatrix}. Let us now rotate $\Sigma$ in the direction of $W$. We can do this by rotating one of basis vectors of $U^{\oplus3}$ (where $\Sigma$ lives), e.g. $e^{1}$, in this direction: $e^{1}\rightarrow e^{1}+\beta\, W$, $\beta\in\mathbb{R}$. Once this rotation has been performed, $\Sigma$ is located in the subspace of $H_{2}(K3,\mathbb{R})$ spanned by
\begin{equation}
  e_{1}, \qquad e^{1}+\beta W, \qquad e_{2}, \qquad e^{2}, \qquad e_{3},  \qquad e^{3}\:.
\end{equation}
For a generic position of $\Sigma$ in this six dimensional space and for generic values of $\beta$, the lattice $\Lambda$ orthogonal to $\Sigma$ is of the type $D_{7}\times E_{8}$.\footnote{In M-theory on the $K3$ surface given by $\Sigma$ this leads to the gauge group $SO(14)\times U(1)\times E_{8}$.} Reinterpreting~\eqref{eq:e8rootsystem} as a set of simple roots of $\Gamma_{E_{8}\times E_{8}}$, this can be understood from the fact that the cycle corresponding to $\alpha_{1}$ as well as the cycle corresponding to the highest root (\ref{eq:e8highestroot}) acquire finite volume.
For $\beta=1$, we find that $\alpha_{9}+e_{1}$ is orthogonal to $e^{1}+\beta W$ (and hence to $\Sigma$) and we therefore have a further 
independent shrinking cycle. This results in a change of singularity type to $D_{8}\times E_{8}$.\footnote{This corresponds to gauge enhancement $SO(16)\times E_{8}$ in M-theory.} The fact that we found another shrinking cycle is due to the integrality of $\alpha_{9}\cdot W$. Thus the orthogonality of $\alpha_{9}+e_{1}$ to $e^1+\beta W$ for $\beta=1$ corresponds to the previously discussed condition for the highest root to survive after introducing the Wilson line $\beta W_I$.

This reasoning can be extended to a generic rotation of $\Sigma$ into the $E_{8}\times E_{8}$ block. In the most general case, three 
vectors $W^i$ are introduced and every orbifold point in $M_{K3}$ can be reached.
All of this is expected, given the well known duality between M-theory compactified on $K3$
and the heterotic string compactified on $T^{3}$ \cite{Aspinwall:1996mn,key-63}.
The $W^i$ are the three Wilson lines that can be used to break the gauge symmetry on the heterotic side.

\section{The \boldmath$T^{4}/\mathbb{Z}_{2}$ orbifold limit of \boldmath$K3$}\label{sec4}

We begin with some definitions regarding $T^{4}/\mathbb{Z}_{2}$. The non-trivial element
of $\mathbb{Z}_{2}$ acts as $-1$ on all the coordinates $x_i$ ($i=1,...,4$) 
of~$T^{4}$. After modding out, the 16 points of $T^{4}$ fixed
under the $\mathbb{Z}_{2}$-action lead to 16 $A_1$~singularities. Their locations are at
\begin{equation}\label{eq:SingLoc}
(x_1,x_2,x_3,x_4)=(\xi_1,\xi_2,\xi_4,\xi_4),\qquad \mbox{with} \qquad \xi_i=0,\frac12 \:.
\end{equation}

The 2-cycles of $T^4$ are all even with respect to $\mathbb{Z}_{2}$ and survive the orbifolding. An integral basis 
is given by the six 2-tori $\pi_{ij}$ corresponding to the $x_i$-$x_j$-plane. Their intersection numbers are 
\begin{equation}
\pi_{ij}\cdot\pi_{ml}=2\varepsilon_{ijml} \:.
\end{equation}
The corresponding Poincar\'e-dual 2-forms are
\begin{equation}
   \mbox{PD}[\pi_{ij}] = \epsilon_{ijpq}\,dx_p\wedge dx_q \:.
\end{equation}
As we will see in more details later, blowing up the 
16 $A_{1}$ singularities of $T^{4}/\mathbb{Z}_{2}$ gives rise to 16 $\mathbb{P}^1$s. They are orthogonal 
with respect to each other and to the torus-cycles $\pi_{ij}$. There is a natural choice of complex structure 
on $T^4/\mathbb{Z}_2$: %compatible with the Enriques involution (i.e. such that $\Omega\mapsto -\Omega$):
$z_1=x_1+\tau_1\,x_4$ and $z_2=x_2+\tau_2\,x_3$ \footnote{
The natural expressions for the K\"ahler form $j$ and the holomorphic two-form $\omega$ are then
$\omega = dz_1 \wedge dz_2$ and $j = a_1 dz_1\wedge d\bar{z}_1 + a_2 dz_2\wedge d\bar{z}_2 + \mbox{Re}[b\,dz_1\wedge\bar{z}_2]$,
where $a_1,a_2\in \mathbb{R}$ and $b\in\mathbb{C}$. In terms of the Poincare-dual of the integral cycles $\pi_{ij}$, we have
\begin{equation}
\omega  = \pi_{34}+\tau_1 \pi_{13}+\tau_2\pi_{42}-\tau_1\tau_2\pi_{12} \qquad
j  = \hat{a}_1  \pi_{23}  + \hat{a}_2 \pi_{14} + \mbox{Re}[b (\pi_{34}+\tau_1 \pi_{13}+\bar{\tau}_2\pi_{42}-\tau_1\bar{\tau}_2\pi_{12})],
\end{equation}
where we defined $\hat{a}_1=-2a_1 \mbox{Im}\tau_1$ and $\hat{a}_2=-2a_2 \mbox{Im}\tau_2$.
}.

It is well known that some $K3$ surfaces, including $T^{4}/\mathbb{Z}_{2}$, allow a fixed-point
free involution $\vartheta$ yielding an Enriques surface\footnote{Nikulin classified all involutions of $K3$
reversing the sign of $\omega$ \cite{key-66} and found that they can be labeled by three integers
$(r,a,\delta)$. Only one involution in this classification, $(10,10,0)\equiv\vartheta$,
has no fixed points.}. The action of $\vartheta$ on $T^{4}/\mathbb{Z}_{2}$ is given by \cite{key-65}
\begin{equation}
\vartheta: \qquad z_1\mapsto -z_1+\frac12,\qquad z_2\mapsto z_2+\frac12 \:.\label{ent4}
\end{equation}
Hence, $\pi_{14}$ and $\pi_{23}$ are even under $\vartheta$, while $\pi_{12}$, $\pi_{34}$, $\pi_{13}$ and $\pi_{42}$
are odd. From \eqref{eq:SingLoc} it is clear that the $A_1$ singularities are interchanged pairwise.
We will use the transformation properties of the cycles of $T^{4}/\mathbb{Z}_{2}$ under
this involution to identify them with specific cycles of the $K3$ lattice.

\begin{figure}[tt]
\begin{centering}
\includegraphics[width=7cm]{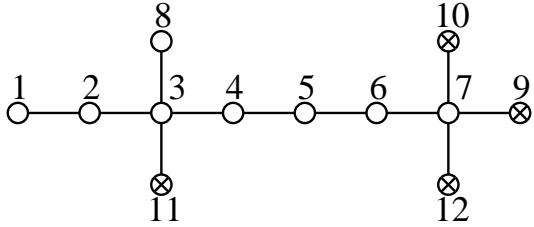}
\par\end{centering}

\caption{The Dynkin diagram of the first $E_{8}$ extended by its highest root as well
as the highest roots of its sublattices of types $D_{4}$ and $D_{8}$.}\label{FigDinkE8ext}

\end{figure}

\
We now discuss the cycles of $T^{4}/\mathbb{Z}_{2}$ from the $K3$ perspective.
The singular limit $T^{4}/\mathbb{Z}_{2}$ of $K3$ is obtained
by fixing the position of $\Sigma$ such that $16$ cycles with intersection
matrix $A_{1}^{\oplus16}$ shrink. We start with a $K3$ surface with
an $E_{8}\times E_{8}$ singularity and rotate $\Sigma$ to a $A_1^{\oplus 16}$ point. %Correspondingly, $\Sigma$ is located inthe $U^{\oplus3}$ block of $H^{2}(K3,\mathbb{R})$.
In Sect.~\ref{sec2} we have specified Wilson lines breaking $E_8$ to $A_1^{8}$. Using these Wilson lines and following the procedure detailed in 
Sect.~\ref{sec3}, we will arrive at the desired point in moduli space.

We introduce a set of simple roots $\gamma_{i}$, $i=1,...,8,13,...,20$, of $\Gamma_{E_{8}\times E_{8}}$
(cf. \eqref{eq:e8rootsystem}), the highest roots $\gamma_{9}$ and $\gamma_{21}$
of the $E_{8}$ root lattices as well as the highest roots $\gamma_{i}$,
$i=10,11,12,22,23,24$, of their respective sublattices of types $D_{4}$ and $D_{8}$ (see Fig.~\ref{FigDinkE8ext}):
\begin{align}\label{eq:e8e8}
\gamma_{1} & =\frac{1}{2}E_{1}+\frac{1}{2}E_{2}+...+\frac{1}{2}E_{8}  &\qquad& \gamma_{2} =-E_{7}-E_{8}  &\qquad& \gamma_{3} =-E_{6}+E_{7} \nonumber \\
\gamma_{4} & =-E_{5}+E_{6}  &\qquad& \gamma_{5} =-E_{4}+E_{5} &\qquad& \gamma_{6} =-E_{3}+E_{4} \nonumber \\
\gamma_{7} & =-E_{2}+E_{3}  &\qquad& \gamma_{8} =-E_{7}+E_{8} &\qquad& \gamma_{9} =-E_{1}+E_{2} \nonumber \\
\gamma_{10} & =E_{1}+E_{2}  &\qquad& \gamma_{11} =E_{5}+E_{6} &\qquad& \gamma_{12} =-E_{3}-E_{4} \nonumber \end{align}
\begin{align}
\gamma_{13} & =\frac{1}{2}E_{9}+\frac{1}{2}E_{10}+...+\frac{1}{2}E_{16} &\qquad& \gamma_{14}  =-E_{15}-E_{16} &\qquad& \gamma_{15} =-E_{14}+E_{15}\nonumber \\
\gamma_{16} & =-E_{13}+E_{14} &\qquad& \gamma_{17} =-E_{12}+E_{13} &\qquad& \gamma_{18} =-E_{11}+E_{12}\nonumber \\
\gamma_{19} & =-E_{10}+E_{11} &\qquad& \gamma_{20} =-E_{15}+E_{16} &\qquad& \gamma_{21} =-E_{9}+E_{10}\nonumber \\
\gamma_{22} & = E_{9}+E_{10}  &\qquad& \gamma_{23} = E_{13}+E_{14} &\qquad& \gamma_{24} =-E_{11}-E_{12}\end{align}

On the basis of \eqref{eq5}, we choose the following Wilson-line vectors in 
$\Gamma_{E_{8}\times E_{8}}$ (The signs between the two $E_8$ factors will be justified in a moment 
by the properties of the $K3$ lattice under the Enriques involution):
\begin{align}\label{WLT4Z2}
W^1=(1,0^7,-1,0^7), \qquad W^2=(0^4,{-\frac12}^4,0^4,{\frac12}^4)\nonumber\\ W^3=(0^2,-\frac12,\frac12,0^2,-\frac12,\frac12,0^2,-\frac12,\frac12,0^2,-\frac12,\frac12)\:.
\end{align}

We start with $\Sigma$ living in the $U^{\oplus 3}$ space spanned by
$ \hat{e}_1$, $\hat{e}^1$, $\hat{e}_2$, $\hat{e}^2$, $\hat{e}_3$, $\hat{e}^3$.
The first step is to move $\Sigma$ in the direction of $W^1$ by the rotation $\hat{e}^{1}\rightarrow \hat{e}^{1}+ W^{1}$. 
The result is a $K3$ surface with $D_{8}\times D_{8}$ singularity. While $\gamma_{1}$, $\gamma_{9}$, $\gamma_{10}$, 
$\gamma_{13}$, $\gamma_{21}$ and $\gamma_{22}$ are blown up, the cycles
\begin{equation}
\gamma_{9}'  \equiv\gamma_{9}-\hat{e}_{1},\qquad
\gamma_{10}'  \equiv\gamma_{10}+\hat{e}_{1},\qquad
\gamma_{21}'  \equiv\gamma_{21}+\hat{e}_{1},\qquad
\gamma_{22}'  \equiv\gamma_{22}-\hat{e}_{1}
\end{equation}
collapse. $\gamma_{9}'$ ($\gamma_{21}'$) is the additional root appearing in the first (second) $E_8$ lattice. 
$\gamma_{10}'$ ($\gamma_{22}'$) are the corresponding highest roots of $D_{8}$.

Next, we rotate $\hat{e}^{2}\rightarrow \hat{e}^{2}+W^{2}$. Since the products of $\gamma_{2}$, $\gamma_{5}$,
$\gamma_{11}$, $\gamma_{14}$, $\gamma_{17}$ and $\gamma_{23}$ with the rotated basis vector $\hat{e}^{2}+W^{2}$
are all non-zero, these cycles acquire finite volume, while $\gamma_{i}$, $i=3,4,6,7,8,12,15,16,18,19,20,24$, 
and $\gamma_{i}'$, $i=9,10,21,22$, remain orthogonal to $\Sigma$. Out of the roots in \eqref{eq:e8e8}, however, 
we can take those that have an integer product with $W^{2}$ and construct the four further shrunk cycles
\begin{align}
\gamma_{2}'  \equiv\gamma_{2}+\hat{e}_{2},\qquad
\gamma_{11}'  \equiv\gamma_{11}-\hat{e}_{2},\qquad
\gamma_{14}'  \equiv\gamma_{14}-\hat{e}_{2},\qquad
\gamma_{23}'  \equiv\gamma_{23}+\hat{e}_{2}\:.
\end{align}
A set of simple roots for the orthogonal lattice $\Lambda$ is given by 
$\lbrace\gamma_{i}\rbrace_{i=3,4,6,7,8,15,16,18,19,20}$ and 
$\lbrace\gamma_{i}'\rbrace_{i=2,9,10,14,21,22}$. The intersection matrix of these 
simple roots is $D_{4}^{\oplus4}$. We therefore obtained a $K3$ surface with a $D_{4}^{4}$ singularity.

Finally, we rotate $\hat{e}^{3}\rightarrow \hat{e}^{3}+ W^{3}$, go to an $A_{1}^{\oplus16}$ point 
in $M_{K3}$. The roots that are removed from $\Lambda$ are $\gamma_{3}$, $\gamma_{6}$, $\gamma_{7}$, 
$\gamma_{8}$, $\gamma_{15}$, $\gamma_{18}$ and $\gamma_{19}$, $\gamma_{20}$, while the new shrinking cycles are
\begin{align}
\gamma_{6}'  \equiv\gamma_{6}+\hat{e}_{3},\qquad
\gamma_{8}'  \equiv\gamma_{8}+\hat{e}_{3},\qquad 
\gamma_{18}'  \equiv\gamma_{18}+\hat{e}_{3},\qquad
\gamma_{20}'  \equiv\gamma_{20}+\hat{e}_{3}.
\end{align}

To sum up, following the procedure outlined in the last section, we found that $K3$ can degenerate to $T^{4}/\mathbb{Z}_{2}$ if
$\Sigma$ is orthogonal to
%The roots in the in the orthogonal complement of $\Upsilon$ are
\begin{align}\label{16shrCycles}
\gamma_{2} & '=-E_{7}-E_{8}+\hat{e}_{2}, & \gamma_{14}' & =-E_{15}-E_{16}-\hat{e}_{2},\nonumber \\
\gamma_{4} & =-E_{5}+E_{6}, & \gamma_{16} & =-E_{13}+E_{14},\nonumber \\
\gamma_{6}' & =-E_{3}+E_{4}+\hat{e}_{3}, & \gamma_{18}' & =-E_{11}+E_{12}+\hat{e}_{3},\nonumber \\
\gamma_{8}' & =-E_{7}+E_{8}+\hat{e}_{3}, & \gamma_{20}' & =-E_{15}+E_{16}+\hat{e}_{3},\nonumber \\
\gamma_{9}' & =-E_{1}+E_{2}-\hat{e}_{1}, & \gamma_{21}' & =-E_{9}+E_{10}+\hat{e}_{1},\nonumber \\
\gamma_{10}' & =E_{1}+E_{2}+\hat{e}_{1}, & \gamma_{22}' & =E_{9}+E_{10}-\hat{e}_{1},\nonumber \\
\gamma_{11}' & =E_{5}+E_{6}-\hat{e}_{2}, & \gamma_{23}' & =E_{13}+E_{14}+\hat{e}_{2},\nonumber \\
\gamma_{12} & =-E_{3}-E_{4} & \gamma_{24} & =-E_{11}-E_{12}.
\end{align}
This set of cycles provides a primitive embedding of the $A_1^{\oplus 16}$ lattice into $\Gamma_{3,19}$.

The lattice orthogonal to the shrunk cycles $\Upsilon$ is given by integral combinations of the 
following six cycles:
\begin{equation}\label{1stBasisUps}
\hat{e}_{1}, \qquad 2(\hat{e}^{1}+ W^{1}), \qquad \hat{e}_{2}, \qquad 2(\hat{e}^{2}+ W^{2}), \qquad \hat{e}_{3}, \qquad 2(\hat{e}^{3}+ W^{3}) \:.
\end{equation}
The 3-plane $\Sigma$ lives in the subspace of $H_{2}(K3,\mathbb{R})$ spanned by these vectors so that the cycles in $\Upsilon$ 
in general have finite size. We want to identify this lattice with the $T^4/\mathbb{Z}_2$ lattice made up of the $\pi_{ij}$. We will use the transformation properties of the torus-cycles $\pi_{ij}$ under $\vartheta$ to identify them with elements of $\Upsilon$. 

Previously we have seen that the Enriques involution must map the singularities of $T^4/\mathbb{Z}_2$ to each other in pairs. We hence expect that the cycles on the left column in \eqref{16shrCycles} are mapped to the cycles on the right one. %Since the involution acts on the $K3$ lattice such that to exchange the two $E_8$ blocks (i.e. $E_I\leftrightarrow E_{I+8}$), the transformation properties of the $\hat{e}_i$ are fixed. 

Up to automorphism of $\Gamma_{3,19}$, the Enriques involution $\vartheta$ acts on the $K3$ lattice by interchanging the two $E_{8}$ 
as well as the two $U$-blocks, and as $-1$ on the remaining $U$-block \cite{peters} (see also \cite{Berglund:1998va}):
\begin{equation}\label{eq:enrinv}
\vartheta:\,e_{1} \mapsto-e_{1} \qquad e^{1} \mapsto-e^{1} \qquad
e_{2} \leftrightarrow e_{3} \qquad e^{2} \leftrightarrow e^{3} \qquad
E_{I} \leftrightarrow E_{I+8}.
\end{equation}

If we set $\hat{e}_i=e_i$ and apply the transformation \eqref{eq:enrinv} to the 16 cycles in \eqref{16shrCycles}, we do not obtain what we expect, i.e. that the 8 cycles in the left column in \eqref{16shrCycles} are mapped to the ones in the right column. To get this result, we need an Enriques involution such that the $\hat{e}_i$ have definite parity. A sensible identification is thus\footnote{One can check that this transformation provides an automorphism of the lattice $\Gamma_{3,19}$.} %Hence the two sets $\{\hat{e}_i,\hat{e}^i\}$ and $\{e_i,e^i\}$ are equivalent.
\begin{equation}
 \hat{e}_1 = e_1 \qquad \hat{e}_2= e_2-e_3 \qquad \hat{e}_3 = e^2+e^3 \qquad \hat{e}^1=e^1 \qquad \hat{e}^2=e^2 \qquad \hat{e}^3 = e_3 \:.
\end{equation}
Hence, the basis \eqref{1stBasisUps} of $\Upsilon$ becomes
\begin{equation}\label{2ndBasisUps}
e_{1}, \qquad 2(e^{1}+ W^{1}), \qquad e_{2}-e_3, \qquad 2(e^{2}+ W^{2}), \qquad e^2+e^{3}, \qquad 2(e_{3}+ W^{3}) \:.
\end{equation}

Note that the set of vectors \eqref{16shrCycles} could be guessed without any reference to a particular choice of Wilson lines by going directly to the Dynkin diagram language. Then, the involution property of the three orthogonal null vectors $\hat{e}_i$ of the $U^{\oplus 3}$-block that we add to rotate the shrinking cycles is determined by requiring the exchange of the two blocks.

We can now choose an integral basis of $\Upsilon$ with the properties of $\pi_{ij}$, i.e. whose elements 
have definite parity under $\vartheta$ and whose intersection matrix is\footnote{
The matrix $U(2)$ is equal to $\left(\begin{array}{cc} 0&2\\2&0\\ \end{array}\right)$.}
$U(2)^{\oplus 3}$:
\begin{equation}\label{eVSpi}\begin{array}{ccc}
e_{1},&\qquad& 2(e^{1}+e_{1}+ W^{1}),\\ e_{2}-e_{3},&\qquad& e^{2}-e^{3}+2(e_{2}-e_{3}+ W^{2}),\\ e^{2}+e^{3}, &\qquad& e_{2}+e_{3}+2(e^{2}+e^{3}+ W^{3})
\end{array}\end{equation}
One can check that this is an integral basis of $\Upsilon$. %(obtained by automorphism of $\Upsilon$ that is not an automorphism of $\Gamma_{3,19}$. ... ). 
The first two lines of \eqref{eVSpi} give two $U(2)$ blocks odd under $\vartheta$, while the last line gives a $U(2)$ block even under $\vartheta$. Therefore, we may identify $\pi_{23}$ and $\pi_{14}$ with the vectors $e^{2}+e^{3}$ and $e_{2}+e_{3}+2(e^{2}+e^{3}+ W^{3})$ of the last $U(2)$ block and $\pi_{12}$, $\pi_{34}$, $\pi_{13}$ and $\pi_{42}$ with the vectors $e_{1}$, $2(e^{1}+e_{1}+ W^{1})$, $e_{2}-e_{3}$ and $e^{2}-e^{3}+2(e_{2}-e_{3}+ W^{2})$ of the first two $U(2)$ blocks.

%\newpage

\section{\boldmath$T^{4}/\mathbb{Z}_{2}$ as a double cover of \boldmath$\mathbb{P}^1\times\mathbb{P}^1$}
\label{sec5}
In this section, we are going to study the connection of
$T^4/\mathbb{Z}_2$ with a smooth $K3$ from a geometric perspective. 
We show how to find the lattice $H_2(K3,\mathbb{Z})$ in a blow-up of~$T^4/\mathbb{Z}_2$. 

It is well-known that one can construct a smooth $K3$ as a double cover
\cite{peters,Sen:1997bp} over $\mathbb{P}^1\times \mathbb{P}^1$, branched 
along a curve of bidegree $(4,4)$:

\begin{equation}
\tilde{y}^{2}=h_{(4,4)}(\tilde{x}_{1},\tilde{x}_{2},\tilde{z}_{1},\tilde{z}_{2}).\label{eq:cyd}
\end{equation}

There are two algebraic cycles, each given by fixing a point on one of the $\mathbb{P}^1$s.
We call the associated Divisors $D_x$ and $D_z$.
The corresponding curves are tori and represent the generic fibres of two different elliptic
fibrations of the $K3$ surface given by \eqref{eq:cyd}. As (\ref{eq:cyd}) gives two values of $y$ for
a generic point on $\mathbb{P}^1\times \mathbb{P}^1$, we find $D_{x}\cdot D_{z}=2$.

\begin{figure}
\begin{centering}
\includegraphics[height=5cm]{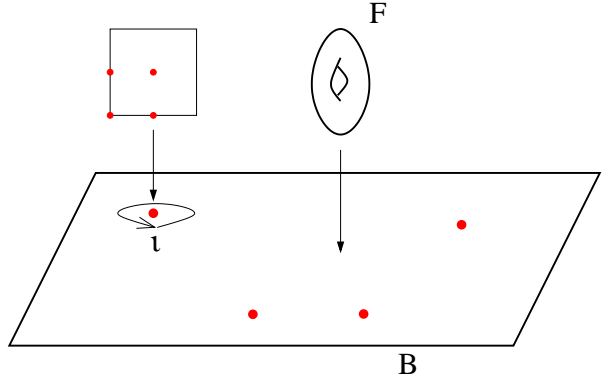}\label{fib}
\par\end{centering}
\caption{The elliptic fibration $T^{4}/\mathbb{Z}_{2}\rightarrow B=T^2/\mathbb{Z}_2$,
has four singular fibres. Upon circling one of them, the fibre torus undergoes an involution 
$\iota$. Thus any section $B\hookrightarrow T^{4}/\mathbb{Z}_{2}$ has to pass through four singularities.}

\end{figure}

Let us choose a particular form for $h_{(4,4)}$:
\begin{equation}
y^{2}=\prod_{k=1,..,4}(x-x_{k})\cdot(z-z_{k}).      \label{eq:alT4}
\end{equation}
For ease of exposition we have introduced the inhomogeneous coordinates
$x,y,z$. The surface defined by \eqref{eq:alT4} is easily recognized as 
$T^4/\mathbb{Z}_2$, as we explain in following: In the vicinity of the points $(y,x,z)=(0,x_{k},z_{h})$, 
it is given by $y^{2}=xz$, i.e. it has sixteen $A_{1}$ singularities.
Let us now describe this surface as an elliptic fibration. We project to the
coordinate $x$, so that each fibre torus is given by (\ref{eq:alT4}) with
$x$ fixed. The complex structure of the fibre torus is given by
the ratios of the branch points $z_k$. As these do not depend on $x$, the complex structure of
the fibre is constant. Over the four points in the base where $x=x_k$, we have $y=0$, so
that the fibre is $\mathbb{P}^1$ instead of $T^2$, see Fig.~\ref{fib}. A
similar fibration is obviously obtained when projecting to the $z$-coordinate.

This is the very same structure one finds when projecting $T^4/\mathbb{Z}_2$ to any
$T^2/\mathbb{Z}_2$ suborbifold. Any of these projections can be promoted to an elliptic fibration
by choosing the complex structure of $T^4/\mathbb{Z}_2$ appropriately. Only two 
of them can, however, be seen algebraically in (\ref{eq:alT4}).
The divisors $D_{x}$ and $D_{z}$ correspond to multisections\footnote{A section is a
divisor that is not part of any fibre and intersects each fibre once. Correspondingly, a multisection or 
$m$-section intersects each fibre $m$ times.} (two-section) of these
two fibrations. They are tori and can be identified with $\pi_{23}$ and $\pi_{14}$. The
other $\pi_{ij}$ in $T^4/\mathbb{Z}_2$ cannot be seen algebraically.

Each of the two elliptic fibrations in (\ref{eq:alT4}) has four proper sections. 
Focussing again on the fibration given by projecting to the $x$-coordinate, they are given 
by mapping $x$ to $(y,x,z)=(0,x,z_k)$. Each of them passes through four $A_1$ singularities.
From the orbifold point of view, these sections are the usual 
divisors ($D_{i\alpha}=\{\zeta_i=\zeta_i^{\alpha,fixed}\}$), given by 
planes lying at the fixed loci of the orbifold action \cite{Lust:2006zh}.

We can understand how these sections arise in $T^4/\mathbb{Z}_2$. Fixing a projection, we have to
give a point in the fibre for every point of the base in a smooth manner. As the fibre undergoes
an involution when one surrounds one of the $x_k$ in the base, the sections have to pass through one
of the fixed points of this involution in the fibre. Again, not all of the sections that can be seen
this way in $T^4/\mathbb{Z}_2$ can be described algebraically in (\ref{eq:alT4}).

We label the sections $\sigma_{ij}^{k}$ by the two directions it spans in $T^4/\mathbb{Z}_2$ ($i,j$) 
and the fixed point in the fibre it passes through ($k$). Two $\sigma_{ij}^{k}$
that span different directions in $T^4/\mathbb{Z}_2$ are, of course, sections with respect
to different elliptic fibrations. The intersection numbers with the $\pi_{ij}$ are
\begin{equation}\label{eq:inT4}
\sigma_{ij}^{k}\cdot\pi_{lm}=\varepsilon_{ijlm}.
\end{equation}
As the intersections occur away from the singularities, \eqref{eq:inT4} will persist in a desingularized version of
$T^4/\mathbb{Z}_2$.

\begin{figure}
\begin{centering}
\includegraphics[height=9cm]{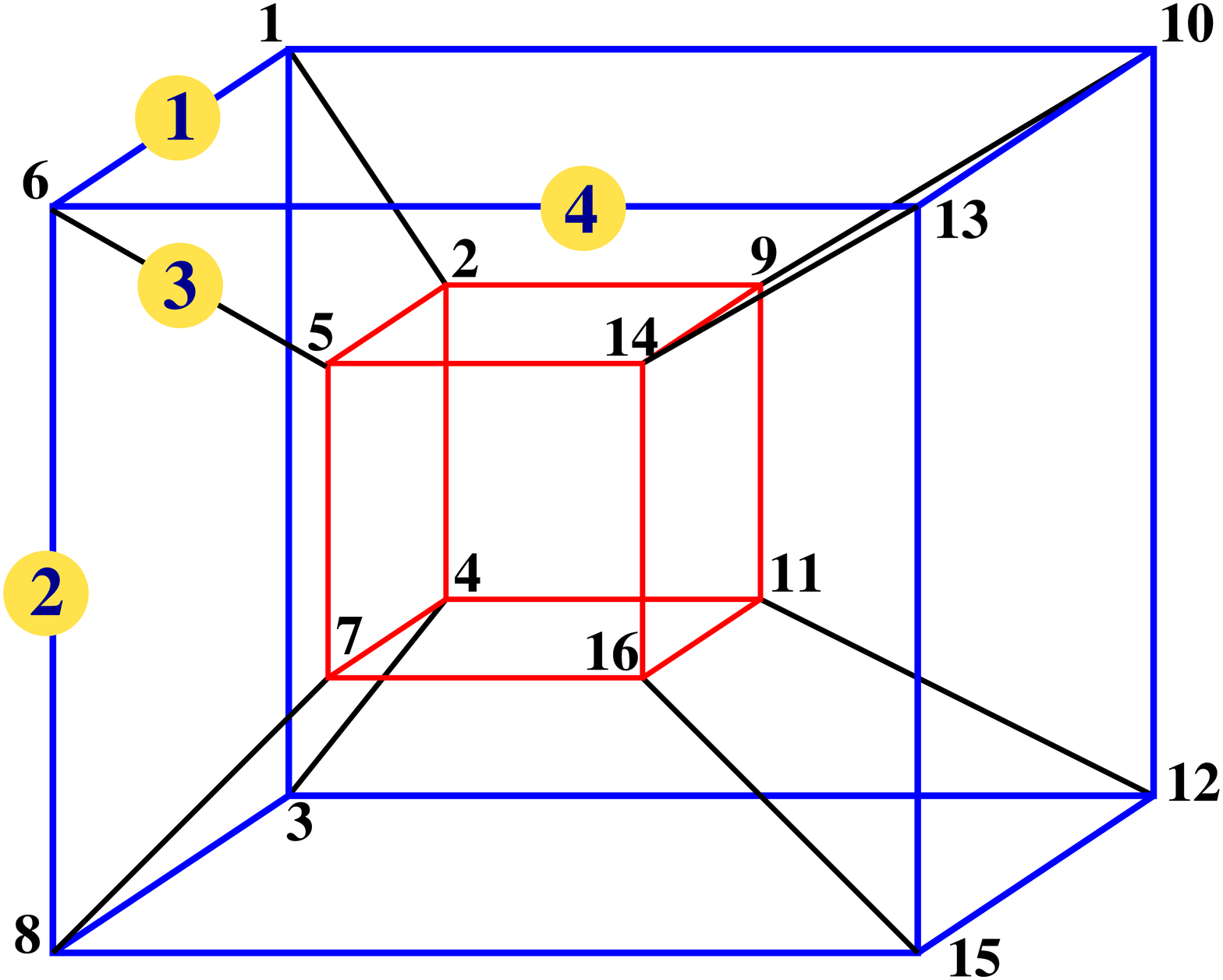}
\caption{The set of sections $\sigma_{ij}^{k}$ and the $A_1$ singularities 
can be displayed as the two-dimensional faces and nodes of a four-dimensional 
hypercube. We picture this cube as two cubes of lower dimension whose nodes are connected 
as shown in the picture. We have numbered the four directions and the sixteen nodes, 
so that this figure can be used to determine which section meets which singularities.}
\label{3cube} 
\end{centering}
\end{figure}

One way to visualize the geometry of $T^4/\mathbb{Z}_2$ is presented in Fig.~\ref{3cube}. It will be
frequently used in the rest of this paper. At present, it serves to determine which singularities are 
met by which $\sigma_{ij}^{k}$ in the given labeling. 

\subsection{Divisors and cycles in the blow-up of $T^{4}/\mathbb{Z}_{2}$}
\label{sec6}

If we blow-up the sixteen $A_{1}$ singularities of $T^{4}/\mathbb{Z}_{2}$,
we introduce sixteen exceptional divisors $C_{\lambda}$ which satisfy $C_{\lambda}\cdot C_{\eta}=-2\delta_{\lambda\eta}$.
Naively, one would guess that the lattice of integral cycles of the blow-up
of $T^{4}/\mathbb{Z}_{2}$ is thus given by $A_{1}^{\oplus16}\oplus U(2)^{\oplus3}$.
But the blow-up of $T^{4}/\mathbb{Z}_{2}$ should be a smooth $K3$ surface, which
has $U^{\oplus3}\oplus(-E_{8})^{\oplus2}$ as its lattice of integral
cycles. The extra integral cycles are given by the preimages of the sections in the 
blow-up\footnote{Remember that a blow-up actually is a projection mapping the blown-up space to the space one starts with 
\cite{GrHarris}.}.

To blow-up an $A_{1}$ singularity (locally given by $y^{2}=xz$
in $\mathbb{C}^{3}$) one introduces an extra $\mathbb{P}^{2}$ with
homogeneous coordinates $\xi_{i}$ and considers the set of
equations (see e.g. \cite{GrHarris})
\begin{equation}
y^{2}  =xz,\qquad \xi_{1}y  =x\xi_{2},\qquad \xi_{1}z  =x\xi_{3}, \qquad \xi_{2}z  =y\xi_{3}
\end{equation}
in $\mathbb{C}^{3}\times\mathbb{P}^{2}$. The exceptional curve $C$ is a $\mathbb{P}^{1}$ given by
\begin{equation}
\xi_{2}^{2}  =\xi_{1}\xi_{3},\qquad\qquad x = y = z = 0.
\end{equation}
Its self-intersection is $C\cdot C=-2$.

The sections $\sigma$ are locally given by $y=x=0$. In the blown up
space $\mathbb{C}^{3}\times\mathbb{P}^{2}$ they are sitting at
\begin{equation}
x=y=0,\qquad\xi_{1}=\xi_{2}=0.%  ,\qquad\xi_{3}=1.
\end{equation}
This shows that the $\sigma_{ij}^{k}$ lead to smooth curves in the blown-up space. 
We furthermore deduce that the $\sigma$ intersect only those exceptional divisors that emerge 
at the four singularities they meet in $T^4/\mathbb{Z}_2$ before the blow-up. The even cycles 
of $T^{4}$, $\pi_{ij}$, are left completely unperturbed by the blow-up and cannot intersect
any of the exceptional divisors. We thus find the following intersections
in the smooth $K3$:
\begin{align}
C_{\lambda}\cdot C_{\eta} & =-2\delta_{\lambda\eta}, &\qquad& \pi_{ij}\cdot\pi_{ml} =2 \varepsilon_{ijml}, &\qquad& C_{\lambda}\cdot\pi_{ij} =0,\\
\sigma_{ij}^{k}\cdot\pi_{ml} & =\varepsilon_{ijml}, &\qquad& \sigma_{jl}^{k}\cdot C_{\lambda} =1\hspace{0.2cm}\mbox{if}\hspace{1ex}i\in I_{jl}^{k}, &\qquad& \sigma_{jl}^{k}\cdot C_{\lambda} =0\hspace{0.2cm}\mbox{if}\hspace{1ex}i\not\in I_{jl}^{k}.\label{eq:inter}\end{align}

The index sets $I_{jl}^{k}$ can e.g. be determined from Fig.~\ref{3cube} 
(remember that the $\sigma_{jl}^k$ correspond to the faces of the hypercube). As we know that the second homology of $K3$ is $22$-dimensional
and the cycles $C_{\lambda}$ and $\pi_{ml}$ are independent, it is clear
that we can use them as a basis for $H_{2}(K3,\mathbb{R})$. Thus
there exists an expansion of the cycles $\sigma_{jl}^{k}$ in terms
of this basis. Using the intersection numbers \eqref{eq:inter}, we 
conclude that\footnote{
This expression is consistent with those of the intersections between the $\sigma_{ij}^{k}$
that can be checked algebraically: If two faces do not meet
at all, their intersection number is clearly zero both algebraically and by \eqref{sigma}.
If they have one node in common, their intersection number is still zero
from (\ref{sigma}). This agrees with the algebraic model where one can check that
these two cycles miss each other in the blown-up $K3$. If two sections have two nodes in 
common, they can not be represented by algebraic subvarieties of \eqref{eq:alT4}. In this 
situation \eqref{sigma} determines their mutual intersection to be unity.}
\begin{equation}
\sigma_{ij}^{k}=\frac{1}{2}\cdot(\pi_{ij}-\sum_{\lambda\in I_{jl}^{k}}C_{\lambda}).\label{sigma}
\end{equation}

Before, we have shown that the $\sigma_{ij}^{k}$ are in fact elements
of the \emph{integral} homology of the smooth, blown-up $K3$. On the other hand we see
from \eqref{sigma} that they are not integral combinations of the $\pi_{ij}$ and $C_\lambda$. 
This tells us that the lattice of integral cycles consists of many more elements than
the ones in $A_{1}^{\oplus 16}\oplus U(2)^{\oplus3}$: it must also contain all
elements of the form \eqref{sigma}. It can moreover be shown that out of the $\sigma_{ij}^k$
and $C_\lambda$ one can construct a basis of integral cycles 
that has an intersection matrix with determinant minus one. 
As all self-intersections are even numbers, we have thus constructed an even 
unimodular lattice of signature $(3,19)$. This lattice must be $\Gamma_{3,19}=U^{\oplus3}\oplus(-E_{8})^{\oplus2}$,
the lattice of integral cycles of $K3$\cite{key-49}. 

Note that the symmetries of $T^{4}/\mathbb{Z}_{2}$ are manifest in our construction. They simply
correspond to a relabeling of or a reflection along one of the four directions of the cube in~Fig.~\ref{3cube}.

A similar construction of $\Gamma_{3,19}=H_2(K3,\mathbb{Z})$ has recently been discussed in 
\cite{Kumar:2009zc}. There it is exploited that $H_2(K3,\mathbb{Z})$ must be an unimodular lattice. 
This property of $H_2(K3,\mathbb{Z})$ is used to systematically 
enlarge $U(2)^{\oplus 3}\oplus A_1^{\oplus 16}$ to $\Gamma_{3,19}$. Our presentation differs in that 
we \emph{geometrically} identify the elements $\sigma_{ij}^k$ that enlarge the lattice
$U(2)^{\oplus 3}\oplus A_1^{\oplus 16}$ to $\Gamma_{3,19}$.

Related discussions of how to find integral cycles after blowing up singularities appear in \cite{Lust:2006zh} 
in the context of type IIB compactifications and in \cite{Nibbelink:2008tv} in the context of heterotic orbifolds (see also \cite{Heter1,Heter2,Heter3,Heter4}).

\subsection{Juxtaposition}
\label{sec7}
In the first part of this section, 
we have given a detailed description of the six finite size cycles of $T^4/\mathbb{Z}_2$ and of its
collapsed cycles. We have then identified them with holomorphic cycles in an algebraic model. Using the results of
Sect.~\ref{sec4}, we are able to match these cycles with the conventionally labelled $K3$ lattice of Sect.~\ref{sec3} \footnote{
This shows that the embedding of $A_1^{\oplus 16}\subset H_2(K3,\mathbb{Z})$ obtained
in the first half of this paper is identical with the embedding of the $C_\lambda$ that is implicit from the last section.}:

The six torus cycles $\pi_{ij}$ are:
\begin{align}
\pi_{23}&=e^2+e^3&\qquad& \pi_{14}=e_2+e_3+2(e^2+e^3)+2W^3\nonumber \\
\pi_{12}&=e_1 &\qquad& \pi_{34}=2(e^1+e_1)+2W^1\nonumber\\ 
\pi_{13}&=e_2-e_3 &\qquad& \pi_{42}=e^2-e^3+2(e_2-e_3)+2W^2\:,
\end{align}
with $W^1$, $W^2$, $W^3$ given in \eqref{WLT4Z2}.

The exceptional cycles $C_\lambda$ are identified with the cycles in (\ref{16shrCycles}):
\begin{align}
C_{1} & =E_{1}+E_{2}+e_1 & C_{9} & =E_{9}+E_{10}-e_{1}\nonumber \\
C_{2} & =-E_{1}+E_{2}-e_1 & C_{10} & =-E_{9}+E_{10}+e_1\nonumber \\
C_{3} & =-E_{3}-E_{4} & C_{11} & =-E_{11}-E_{12} \nonumber \\
C_{4} & =-E_{3}+E_{4}+e^{2}+e^{3} & C_{12}& =-E_{11}+E_{12}+e^{2}+e^{3}\nonumber \\
C_{5}& =-E_{5}+E_{6} & C_{13} & =-E_{13}+E_{14}\nonumber \\
C_{6} & =E_{5}+E_{6}-e_{2}+e_{3} & C_{14}& =E_{13}+E_{14}+e_{2}-e_{3}\nonumber \\
C_{7} & =-E_{7}+E_{8}+e^{2}+e^{3} & C_{15} & =-E_{15}+E_{16}+e^2+e^3\nonumber \\
C_{8}& =-E_{7}-E_{8}+e_2-e_3 & C_{16} & =-E_{15}-E_{16}-e_2+e_3\:.\label{Ci}
\end{align}
% By writing down these assignments we have of course assumed that the
% two embeddings are in fact the same. What remains to be checked is the
% consistency of this assumption: both sides of the equations should 
% declare the same vectors to be lattice points in $U^3\oplus -E_8^{\oplus 2}$.
A non-trivial check of the identifications made above is to use \eqref{intk3} to show 
that all of the $\sigma_{ij}^k$ as given in (\ref{sigma}) are indeed elements
of the $K3$ lattice. The results are collected in the appendix.

We can now easily write down 
the roots of $E_8 \times E_8$ and the basis vectors $e_i, e^i$ of the three 
hyperbolic lattices in terms of the integral cycles we have found in the blow-up.
In terms of the standard labeling, they are given by
\begin{align}
&1:\frac{1}{2}\sum_{i=1}^8 E_i  =-\sigma_{12}^1-C_{3}-C_{8}+\pi_{13} &\qquad &2: -E_7-E_8  = C_{8}-\pi_{13} \nonumber \\
&3: -E_{6}+E_{7}  = \sigma_{23}^1 &\qquad &4:-E_{5}+E_{6}  = C_{5} \nonumber \\
&5:-E_{4}+E_{5}  = \sigma_{13}^1-\sigma_{23}^2+\pi_{23}+C_{6}-C_{4} &\qquad &6:-E_{3}+E_{4}= C_{4}-\pi_{23}\nonumber \\
&7:-E_{2}+E_{3}=\sigma_{23}^2 &\qquad &8:-E_{7}+E_{8}=C_{7}-\pi_{23}
\end{align}
for the first $E_8$ and by
\begin{align}
&1:\frac{1}{2}\sum_{i=9}^{16} E_i  =-\sigma_{12}^3-C_{16}-C_{11}+\pi_{12}-\pi_{13} &\qquad &2:-E_{15}-E_{16}=C_{16}+\pi_{13} \nonumber \\  
&3:-E_{14}+E_{15}=\sigma_{23}^4 &\qquad &4:-E_{13}+E_{14}=C_{13} \nonumber \\
&5:-E_{12}+E_{13}= \sigma_{13}^2-\pi_{13}-\sigma_{23}^3+\pi_{23}+C_{14}-C_{12} &\qquad &6:-E_{11}+E_{12}= C_{12}-\pi_{23}\nonumber \\
&7:-E_{10}+E_{11}=\sigma_{23}^3 &\qquad &8:-E_{15}+E_{16}=C_{15}-\pi_{23}
\end{align}
for the second $E_8$. We furthermore find that
\begin{align}
e_1&=\pi_{12} &\qquad& e^1=\sigma_{34}^2+C_{2}+C_{9}+\pi_{12} \nonumber \\
e_2&=\pi_{13}+e_3 &\qquad& e^2=\pi_{23}-e^3 \\
e_3&=\sigma_{14}^1-C_{7}-C_{4}-\sigma_{13}^4+\pi_{23}  &\qquad& e^3=\sigma_{42}^2-C_{16}+C_{8}+\sigma_{23}^1-\pi_{13}\:.\nonumber
\end{align}

\section{The Enriques involution}
\label{sec8}
In this section we will describe the Enriques involution in detail.

Let us first determine its action on the sixteen $A_1$ singularities and the corresponding exceptional 
divisors from its action on $H_2(K3,\mathbb{Z})$ 
\cite{peters,Berglund:1998va}:
\begin{equation}
\vartheta:\,e_{1} \mapsto-e_{1} \qquad e^{1} \mapsto-e^{1} \qquad
e_{2} \leftrightarrow e_{3} \qquad e^{2} \leftrightarrow e^{3} \qquad
E_{I} \leftrightarrow E_{I+8}\:. \label{actenr}
\end{equation}
From \eqref{Ci} we see that $C_\lambda\leftrightarrow C_{\lambda+8}$. Considering Fig.~\ref{3cube}, this
means that the singularities are exchanged along the $3$-$4$-directions. This can be reproduced from the action 
of the Enriques involution on $T^4/\mathbb{Z}_2$, see \eqref{ent4}. We can also see the same behavior in the 
description of $T^4/\mathbb{Z}_2$ as a hypersurface, (\ref{eq:alT4}): 
By shifting and rescaling $x$ and $z$, we can always arrange that $x_1=-x_2$, $x_3=-x_4$ and $z_1=-z_2$, $z_3=-z_4$. 
The Enriques involution then acts as $\vartheta:(y,x,z) \mapsto (-y,-x,-z)$ \cite{peters}, so that the sixteen~$A_1$
singularities are exchanged as noted before.

\begin{figure}
\begin{center}
\includegraphics[height=4cm]{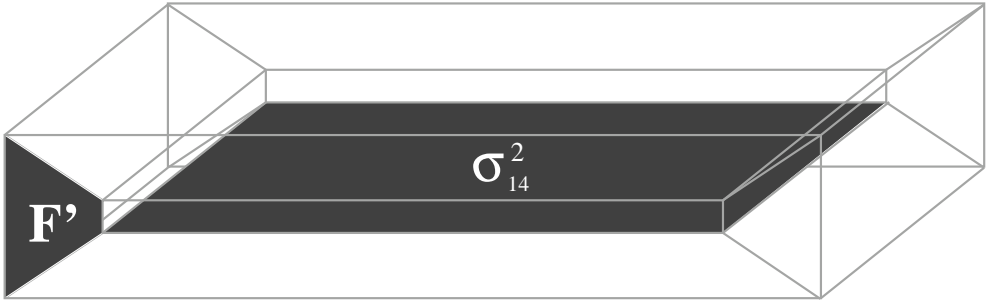}\caption{Choosing the generic fibre to be in the homology class $\pi_{23}$, the
sections are in the $1$-$4$ direction (compare with Fig.~\ref{3cube}). We have depicted the section $\sigma_{14}^2$
and the finite-size component of one of the singular fibres, $F'=\sigma_{23}^1$.}\label{FandB}
\end{center}
\end{figure}

To fix an elliptic fibration of $T^4/\mathbb{Z}_2$, we first select $\pi_{23}=e^2+e^3$ as the homology class of the 
generic fibre. It is obviously invariant under the Enriques involution \eqref{actenr}. The sections are then given
by the $\sigma_{14}^k$, see Figs. \ref{3cube} and \ref{FandB}. 
They can be expressed in terms of the $K3$ lattice as
\begin{align}
\sigma_{14}^1 & =e_3+E_4+E_8\nonumber \\
\sigma_{14}^2 & =e_2+E_{12}+E_{16}\nonumber \\
\sigma_{14}^3 & =e_1+e_3+e^2+e^3 +\frac{1}{2}\left(E_1-E_2-E_3+E_4+E_5-E_6-E_7+E_8\right) \nonumber \\
& +\frac{1}{2}\left(-E_9-E_{10}-E_{11}+E_{12}-E_{13}-E_{14}-E_{15}+E_{16}\right) \nonumber \\
\sigma_{14}^4 & =-e_1+e_2+e^3+e^2 +\frac{1}{2}\left(-E_1-E_2-E_3+E_4-E_5-E_6-E_7+E_8\right) \nonumber \\
& +\frac{1}{2}\left(E_9-E_{10}-E_{11}+E_{12}+E_{13}-E_{14}-E_{15}+E_{16}\right).\label{sectenr}
\end{align}
The Enriques involution acts by exchanging them pairwise. This implies that the resulting Enriques surface is 
elliptically fibred with a two-section, i.e. $\tilde{B}\cdot \tilde{F}=2$. This result is expected from the 
general theory of Enriques surfaces \cite{peters}.
Note that the pairwise exchange of the sections under the Enriques involution can also be seen from (\ref{eq:alT4}).

\subsection{The standard Weierstrass model}\label{sec61}

We now want to make contact with a Weierstrass model with constant $\tau$. It takes the form \cite{Sen:1996vd}
\begin{equation}\label{weier}
y^2=x^3+\alpha_1 h^2xz^4+\alpha_2 h^3z^6=(x-\gamma_1 z^2h)(x-\gamma_2 z^2h)(x-\gamma_3 z^2h).
\end{equation} 
Here $\gamma_i$ and $\alpha_i$ are complex constants and $h$ is a homogeneous polynomial of the base coordinates of degree $4$.
Contrary to $T^4/\mathbb{Z}_2$, the surface described by this equation has four $D_4$ singularities.
There are three sections given by $y=0, x=\gamma_iz^2h$ that pass through the four $D_4$ singularities at $y=x=h=0$. The fourth 
section at $x^3=y^2, z=0$ does not hit any singularity. This section is a special feature of the Weierstrass model and we will 
denote it by $\hat{\sigma}$ in the following.

From what we have said, it is clear that $T^4/\mathbb{Z}_2$ cannot be described by a Weierstrass model. In fact, the section $\hat{\sigma}$ must be orthogonal to all shrinking cycles, and then, for $T^4/\mathbb{Z}_2$ it should belong to $\Upsilon$ (see \eqref{eVSpi}). But this is not possible, since this is a lattice with all intersection numbers being even, and the section $\hat{\sigma}$ should have intersection one with the fibre. %no elliptic fibre and holomorphic section exist whose mutual intersection is one. Thus there can not be a Weierstrass model description of this space.

\begin{figure}[tt]
\begin{center}
\includegraphics[height=55mm]{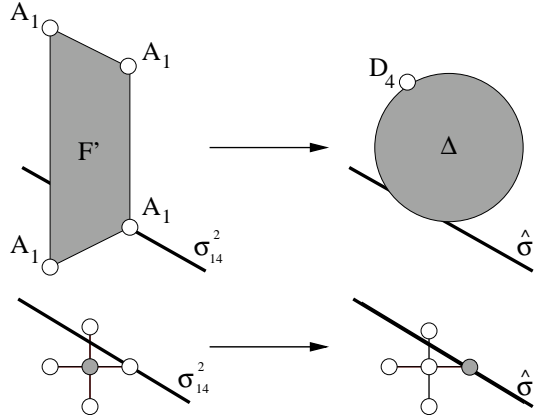}\caption{When blowing up the $A_1$ singularity hit by the section $\sigma^2_{14}$ while collapsing the $F'$ component of the singular fibre, we produce a $D_4$ singularity. This $D_4$ singularity is not hit by the section $\sigma^2_{14}$, which is then identified with 
$\hat{\sigma}$. In this figure we display singularities and collapsed cycles in white and cycles of finite size in light grey. In the lower part of the figure we have drawn the intersection pattern between the cycles in a diagrammatic fashion. After the $F'$ component of the singular fibre is blown down and the $A_1$ singularity hit by the section is blown up, the collapsed cycles intersect according to the Dynkin diagram of $SO(8)$, so that this operation produces a $D_4$ singularity.}\label{a1tod4}
\end{center}
\end{figure}

Intuitively, there is an obvious way how to get from $T^4/\mathbb{Z}_2$ to an elliptic $K3$ described by \eqref{weier}. First,
we choose one of the sections of $T^4/\mathbb{Z}_2$, say $\sigma_{14}^2$, that is to become $\hat{\sigma}$. We then blow up the 
singularities which are hit by this section while blowing down the finite-size components $F'$ of the four singular fibres. We 
have depicted this deformation in Fig.~\ref{a1tod4}. 
The section $\sigma_{14}^2$, which is now identified with $\hat{\sigma}$, no longer intersects any singularities and the lattice 
of collapsed cycles is exactly $D_4^{\oplus 4}$. The other three sections are all forced to meet at the $D_4$ singularities.

After deforming $T^4/\mathbb{Z}_2$ to an elliptic $K3$ described by \eqref{weier}, the $\sigma_{14}^k$ remain sections
of the elliptic fibration. The symmetry among them that is present in $T^4/\mathbb{Z}_2$, however, is lost. This is what 
prevents the Enriques involution from acting on an elliptic $K3$ described by a Weierstrass model like \eqref{weier}.

We can make this more precise using our description of $T^4/\mathbb{Z}_2$ as a point in the moduli space of $K3$, that is, 
the position of the 3-plane $\Sigma$ with respect to the $K3$ lattice of integral cycles.
The prescription that comes from the previous consideration is that one has to move the plane $\Sigma$ such that the cycles intersecting the section $\sigma^2_{14}$ blow up, while the corresponding $F'_k=\sigma_{23}^k$ shrink to zero size.

The four sets of cycles that intersect as in Fig.~\ref{a1tod4} are:
\begin{eqnarray}
 &&C_1,C_2,C_3,C_4,F'_2\equiv\sigma_{23}^2=-E_2+E_3 \qquad \qquad C_5,C_6,C_7,C_8,F'_1\equiv \sigma_{23}^1=-E_6+E_7 \nonumber\\
 &&C_9,C_{10},C_{11},C_{12},F'_3\equiv \sigma_{23}^3=-E_{10}+E_{11} \qquad C_{13},C_{14},C_{15},C_{16},F'_4\equiv \sigma_{23}^4=-E_{14}+E_{15}\nonumber
\end{eqnarray}
Before rotating $\Sigma$, the cycles $C_\lambda$ are shrunk, while the $F'_k=\sigma_{23}^k$ have finite size. To go to the $D_4^{\oplus 4}$ point described by the Weierstrass model, the four cycles $C_4,C_7,C_{11},C_{16}$ must blow up, while the $\sigma_{23}^k$ must shrink. This requires the plane to be located in the subspace orthogonal to $\sigma_{23}^k$ ($k=1,...,4$) and $C_{\lambda}$ ($\lambda=1,2,3,5,6,8,9,10,12,13,14,15$). The latter is generated by:
\begin{equation}\label{Thlatt}\begin{array}{lcl}
\pi_{23}=e^2+e^3,&\qquad& \sigma^2_{14}=e_2+E_{12}+E_{16},\\ \pi_{12}=e_{1},&\qquad& \pi_{34}=2(e^1+e_1+W^1),\\ 
\pi_{13}=e_2-e_3, &\qquad& \pi_{42}=e^2-e^3+2(e_2-e_3+W^2)\:.
\end{array}\end{equation}
These integral cycles generate the lattice contained in this subspace and have intersection matrix\footnote{The corresponding embedding of the $D_4^{\oplus 4}$ lattice in the $K3$ lattice is equivalent (i.e. connected by an automorphism of the $K3$ lattice) to the one given in \cite{key-4}.}:
\begin{equation}
\left(\begin{array}{cccccc}
 0 & 1 & & & & \\ 1 & -2 & & & & \\ & & 0 & 2 & & \\ & & 2 & 0 & & \\ & & & & 0 & 2\\ & & & & 2 & 0\\
\end{array}\right)\:.
\end{equation}
To specify a complex structure compatible with the Enriques involution, we have to choose $\omega$ as a linear combination of the odd cycles in \eqref{Thlatt} and $j$ as a linear combination of the even cycles in \eqref{Thlatt}. As $\pi_{23}$ is the only even cycle, $j$ must be proportional to it. This, however, violates the requirement $j\cdot j>0$.

\

We can also see the clash between the Weierstrass model description and the Enriques involution from a different perspective.
We start with the 3-plane $\Sigma$ in the lattice $\Upsilon$ (see \eqref{eVSpi}). %Since this is a lattice with all intersection numbers being even, no elliptic fibre and holomorphic section orthogonal to the shrinking $A_1$-cycles exist whose mutual intersection is one. Thus there can not be a Weierstrass model description of this space.
Since we want a complex structure compatible with a holomorphic Enriques involution, we take $\omega$ in the odd subspace of $\Upsilon$.
We now try to make the rotation to an $SO(8)^4$ point, maintaining the symmetry under the Enriques involution.
Since the third Wilson line $W^3$ is symmetric under the 
Enriques involution, we can switch it off without destroying the symmetry. Note that this means that we have only changed $j$.
From the discussion of Sect.~\ref{sec2} to Sect.~\ref{sec4} it is clear that removing $W^3$ will result in a $K3$ with four $D_4$ singularities. 
Now the 3-plane $\Sigma$ lives in\footnote{Notice that, in contrast to \eqref{Thlatt}, these cycles do not generate the lattice orthogonal to the shrinking cycles.}
\begin{equation}\label{Thlatt2}\begin{array}{lcl}
\pi_{23}=e^2+e^3,&\qquad& e_2+e_3,\\ \pi_{12}=e_{1},&\qquad& \pi_{34}=2(e^1+e_1+W^1),\\ 
\pi_{13}=e_2-e_3, &\qquad& \pi_{42}=e^2-e^3+2(e_2-e_3+W^2)\:.
\end{array}\end{equation}
Comparing with \eqref{Thlatt}, we see that we have replaced $\sigma^2_{14}$ by $e_2+e_3$. This means that this time we have blown up the cycles $C_4,C_7,C_{12},C_{15}$ while shrinking the cycles $\sigma^k_{23}$. %Looking back at Fig.~\ref{3cube} one can check that there is no holomorphic section $\hat{\sigma}$ that hits none of the singularities.
When $\Sigma$ lives in \eqref{Thlatt2}, we can find a section that does not meet any singularities, e.g. $e_2-(e^2+e^3)$. However, there exists no choice for $\omega$ such that $\omega$ is odd and orthogonal to this section at the same time. In fact, these two conditions require $\omega$ to live in a subspace with degenerate metric, as can be seen by looking back at \eqref{Thlatt2}. 
The complex structure that is demanded by the holomorphicity of the section $\hat{\sigma}$ and the complex structure demanded by the Enriques involution are not compatible.

In summary: Starting from $T^4/\mathbb{Z}_2$, there are two ways to rotate the 3-plane $\Sigma$ such as to get a Weierstrass model with 
$D_4^{\oplus 4}$ singularity. They have different behavior with respect to the Enriques involution: In the first case, we destroy the symmetry. 
In the second case, the symmetry is preserved, but there is no choice of complex structure that both admits a holomorphic section (which does not hit 
the singularities) and makes the Enriques involution holomorphic.

\subsection{A symmetric Weierstrass model}\label{sec62}

In the standard Weierstrass model description, in which the fibre is embedded as a hypersurface in  $\mathbb{P}_{1,2,3}$,
one always has one section $\hat{\sigma}$. By embedding the fibre in other spaces, it is
possible to obtain elliptic $K3$ surfaces with two or more sections \cite{Klemm:1996ts,Berglund:1998va}. 
In particular, it is known that embedding the fibre in $\mathbb{P}_{1,1,2}$ yields an elliptic $K3$ with
two sections which are permuted under the Enriques involution \cite{Berglund:1998va}. 
This elliptic $K3$ is given by an equation of the form
\begin{equation}
y^2=x^4+x^2z^2f_4+z^4f_8.\label{2weier}
\end{equation}
The $\mathbb{Z}_2$ transformation $(y,x,z)\mapsto (-y,x,-z)$ together with a holomorphic involution of the $\mathbb{P}^1$ base
has no fixed points and projects out the holomorphic two-form, so that it provides an Enriques involution of $K3$.
The two holomorphic sections $\hat{\sigma}_1,\hat{\sigma}_2$ are given by $z=0, y=\pm x^2$ and are permuted 
under the Enriques involution. The j-function of this fibration is given by \cite{Berglund:1998va}:
\begin{equation}
\frac{1}{108}\frac{(f_4^2+12f_8)^3}{f_8(-f_4^2+4f_8)^2}.
\end{equation}
Let us discuss the limit in which the complex structure of the fibre is constant. To achieve this, we
take $f_8=f_4^2$. Setting $z=1$ and shifting $f_4$ by some multiple of $x^2$ to complete the square\footnote{Note that 
this is a bijective map between the coordinates $y,x,f_4$ and $y,x,f_4'$.}, we find the equation 
\begin{equation}
y^2=f_4'^2+x^4.
\end{equation}
Thus there are four $A_3$ singularities at the four points $f_4=x=y=0$. 

\begin{figure}
\begin{center}
\includegraphics[height=6cm]{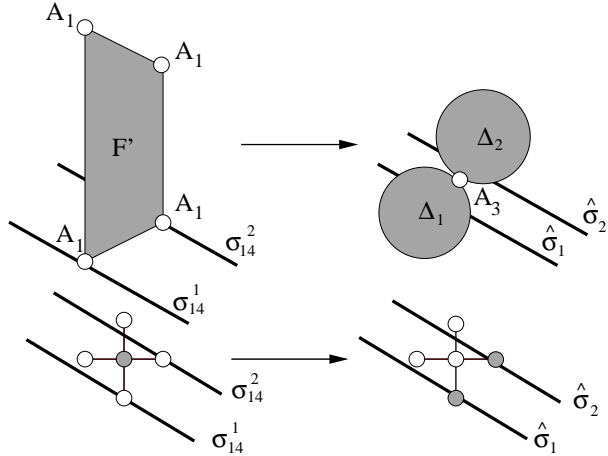}\caption{When blowing up two $A_1$ singularities while shrinking the 
finite-size component of the singular fibre, we produce an $A_3$ singularity and two sections, $\hat{\sigma}_1$ and $\hat{\sigma}_2$, 
which do not hit any singularities. This works in a similar way as the deformation of $T^4/\mathbb{Z}_2$ to an $SO(8)^{\oplus 4}$
configuration, see Fig.~\ref{a1tod4}. We again display singularities and collapsed cycles in white and cycles 
of finite size in light grey.}\label{a1toa3}
\end{center}
\end{figure}

Let us find this configuration by deforming $T^4/\mathbb{Z}_2$. The strategy is similar to that employed
for the deformation of $T^4/\mathbb{Z}_2$ to a $D_4^{\oplus 4}$ configuration. In order to get two sections that
do not hit any singularities and that are interchanged by the Enriques involution we have to blow up
$C_\lambda$, $\lambda=3,4,7,8,11,12,15,16$. At the same time we shrink the cycles $\sigma_{23}^k$ to produce four $A_3$ singularities,
see Fig.~\ref{a1toa3}.

This can be realized by\footnote{We have chosen $j$ and $\omega$ in a six-dimensional subspace of the 10-dimensional space
orthogonal to the~12~$A_3^{\oplus 4}$ cycles. The lattice of cycles orthogonal to a generic $\Sigma$, i.e. orthogonal to the six
basis cycles of \eqref{a34}, then has a dimension 
which is bigger than~12. By examining this lattice, one can check that in spite 
of this the singularity is still $A_3^{\oplus 4}$. Another way to see this is through the associated 
Wilson-line breaking.}%: the Wilson-line appearing in $j$ ($W^{3'}=(0^3,\frac12,0^3,\frac12,0^3,\frac12,0^3,\frac12)$) breaks $SO(8)$ without preserving its rank.}Maximal subgroup??? 
%Rob give Wilson-line more on this kind of breaking ?
\begin{eqnarray}
j&=&b\,\pi_{23}+f\pi_{14}-\frac{f}{2}\sum_\lambda C_\lambda=b\,\pi_{23}+f\left(\sigma_{14}^1+\sigma_{14}^2\right), \nonumber \\
\omega&=&\pi_{34}+U\,\pi_{13}+S\,\pi_{42}-U\,S\,\pi_{12}.   \label{a34}
%\omega&=&s_1\pi_{12}+s_2\pi_{34}+s_3\pi_{13}+s_4\pi_{42}.
\end{eqnarray}
Here $f$ gives the volume of the elliptic fibre. 
The two sections $\hat{\sigma}_{1}=\sigma_{14}^1$ and $\hat{\sigma}_{2}=\sigma_{14}^2$ are orthogonal to $\omega$ and do not intersect any
of the collapsed cycles. $j$ and $\omega$ have the right transformation properties under the Enriques involution.

\section{F-theory Limit}\label{sec9}

There is more than one way to construct an elliptic Calabi-Yau (n+1)-fold that describes a type IIB orientifold compactification 
on a Calabi-Yau n-fold $CY_n$ with D7-charge cancelled locally\footnote{Running Sen's weak coupling limit \cite{Sen:1996vd,Sen:1997gv,Sen:1997kw} backwards, a general procedure to construct an F-theory Calabi-Yau 4-fold, given a generic type IIB setup with D7-branes and O7/O3-planes, was obtained in \cite{Collinucci:2008zs,Blumenhagen:2009up}.}. The examples we discuss here fall into two classes:
\begin{enumerate}
\item Weierstrass models with constant $\tau$.
\item Fourfolds $(CY_n\times T^2)/\mathbb{Z}_2$, where the $\mathbb{Z}_2$ acts as an orientifold 
involution on $CY_n$ and inverts the complex coordinate $z$ of $T^2$, see e.g \cite{key-68}.
\end{enumerate}
The corresponding M-theory backgrounds are different. It is only in the F-theory limit that 
they are dual to the same type IIB background. We will illustrate this fact for the simple examples described in this paper
and consider an elliptically fibred Calabi-Yau two-fold, i.e. $K3$, whose fibre has a constant complex structure. 
We consider two different types of Weierstrass models with constant $\tau$ and the $T^4/\mathbb{Z}_2$ limit of $K3$:

\begin{itemize}
\item \textit{An elliptically fibred $K3$ with one distinguished section.} There are four points on the base $\mathbb{P}^1$ where 
the 2-fold develops a $D_4$ singularity. The cycles corresponding to fibre and section are $F=\pi_{23}$ and 
$\hat{\sigma}=\sigma_{14}^2$. The K\"ahler form and the complex structure live in the space \eqref{Thlatt}. They can be chosen as\footnote{The most general expression for $j$ also includes two deformations in $\langle \pi_{12},\pi_{34},\pi_{13},\pi_{42}\rangle$; we do not include these here, as they are not relevant for the 7-dimensional gauge group and go to zero in the F-theory limit \cite{Braun:2008pz,rob}.}:
\begin{eqnarray}
j&=&b\,\pi_{23}+f\sigma_{14}^2, \nonumber \\
\omega&=&\pi_{34}+U\,\pi_{13}+S\,\pi_{42}-U\,S\,\pi_{12}.
\end{eqnarray}
This point in moduli space is the one reached from $T^4/\mathbb{Z}_2$ by the rotation of $j$ described in Sect.~\ref{sec61}.
\item \textit{An elliptically fibred $K3$ with two distinguished sections.} Again there are four points on the base where
the two-fold develops a singularity. This time, however, this is an $A_3$ singularity, as described in Sect.~\ref{sec62}.
The K\"ahler form and the complex structure can be given by
\begin{eqnarray}
j&=&b\,\pi_{23}+f\left(\sigma_{14}^1+\sigma_{14}^2\right), \nonumber \\
\omega&=&\pi_{34}+U\,\pi_{13}+S\,\pi_{42}-U\,S\,\pi_{12}.
\end{eqnarray}
\item \textit{The space $(T^2\times T^2)/\mathbb{Z}_2$, i.e. the $T^4/\mathbb{Z}_2$ limit of $K3$.} This manifold has 
16 $A_1$ singularities. One choice for the K\"ahler form and the complex structure is
\begin{eqnarray}
j&=&b\,\pi_{23}+f\pi_{14}, \nonumber \\
\omega&=&\pi_{34}+U\,\pi_{13}+S\,\pi_{42}-U\,S\,\pi_{12}.
\end{eqnarray}
\end{itemize}
We notice that the only difference between the three cases is the expression for the K\"ahler form $j$.

Compactifying M-theory on these manifolds gives different 7-dimensional spectra, as all three have different singularities. In particular, we obtain the gauge groups $SO(8)^4$ in the first case, $SO(6)^4\times U(1)^4$ in the second case and $SU(2)^{16}$ in the third case. In the dual type IIB model on $S^1_B\times T^2/\mathbb{Z}_2$, we have four D7-branes on top of each O7-plane wrapping $S^1_B$. However, in the second and third case the gauge group is broken by Wilson lines along the~$S^1_B$~\footnote{The deformations of $j$ inside the $SO(8)$ cycles are mapped to the 8th component of the type IIB vectors (see \cite{rob} for details).}. On the type IIB side, the F-theory limit is given by $R_B\rightarrow \infty$. 
In this limit the $S^1_B$ decompactifies and the Wilson lines become trivial, leaving $SO(8)^4$ as the 8-dimensional gauge group.

\

Let us have a more detailed look at the F-theory limit from the M-theory perspective, i.e., we send the fibre size to zero and see how the 
K\"ahler form and the complex structure behave.

\begin{itemize}
 \item In the first case the F-theory limit is described in \cite{Braun:2008pz}: since the fibre $F$ is orthogonal to $\omega$, 
its size is given by
\begin{equation}
\rho(F) = \int_F j = F\cdot j = f.
\end{equation}
This vanishes in the F-theory limit $f\rightarrow 0$, and the K\"ahler form becomes $j\rightarrow b\,\pi_{23}$.

Note that we find some further shrinking cycles in this limit: $C_4,C_7,C_{11},C_{16}$ only have a finite size due to their intersection with
$\sigma_{14}^2$ in $j$. Letting $f\rightarrow 0$ they collapse so that the intersection pattern of shrunk cycles is now
four times the \emph{extended} Dynkin diagram of $SO(8)$. This is expected from a general perspective: The component of the fibre that
has finite size and the four associated collapsed cycles have the extended Dynkin diagram of $SO(8)$ as their intersection pattern (see Fig.~\ref{a1tod4}). Their sum, i.e the singular fibre, is homologous to the generic fibre, see e.g. \eqref{sigma}. Sending the volume of the 
generic fibre to zero, all five cycles have to collapse. 

\item The second case differs only through the term proportional to $f$ in the K\"ahler form~$j$. In the
limit~$f\rightarrow 0$ we thus reach the same point in the moduli space of $K3$.

\item The same happens for $T^4/\mathbb{Z}_2$. Our choice of $j$ and $\omega$ has of course been completely arbitrary.
Using Fig.~\ref{3cube}, we can easily discuss the most general case: $j$ is then given as
\begin{equation}\label{KaelerFormT4Z2}
 j=f\pi_{ij}+b\,\pi_{ml},
\end{equation}
with four different indices $i,j,m,l$. The holomorphic two-form $\omega$ lives in the space spanned by the
$\pi_{pq}$ that have zero intersection with the K\"ahler form \eqref{KaelerFormT4Z2}. Besides the sixteen cycles $C_\lambda$ we find that
all of the four $\sigma_{ml}^k$ (with $k=1,...,4$) are collapsed when $f\rightarrow 0$. Theses 20 cycles
intersect precisely according to the extended Dynkin diagram of $SO(8)^{\oplus 4}$, as expected. 
\end{itemize}

We have found a geometric realization of the result in \cite{Braun:2008pz}: deforming $j$ does not alter the point in
moduli space reached in the F-theory limit: It sets to zero all components of $j$ except the fibre. If we have multiple
sections, they collapse to a single one in the F-theory limit.

We also found an important result: Before the F-theory limit, the Enriques involution is consistent only with $T^4/\mathbb{Z}_2$ and the symmetric Weierstrass model. {\it In the F-theory limit the Enriques involution is also consistent with the standard Weierstrass model.}

\section{Conclusions and Outlook}

In this note we have obtained an explicit embedding of the lattice of cycles 
that are collapsed when $K3$ degenerates to $T^4/\mathbb{Z}_2$ in $H_2(K3,\mathbb{Z})$.
This embedding leads to a highly symmetric representation of $H_2(K3,\mathbb{Z})$.

We have shown geometrically that $T^4/\mathbb{Z}_2$ and the two Weierstrass-model $K3$s discussed in this paper 
are equivalent in the F-theory limit. Away from that limit, their crucial difference lies in the structure of their sections.
In the case of an elliptic $K3$ described by a standard Weierstrass model, the presence of the distinguished section $\hat{\sigma}$
is inconsistent with the Enriques involution (except in the F-theory limit).

As an application of our detailed description of the Enriques
involution we envisage the generalization of the flux-stabilization
analyses on $K3\times K3$~\cite{gkt04,lmr05,drs99,ak05,Braun:2008pz} to $(K3\times K3)/\mathbb{Z}^E_2$.
Here we assume that $\mathbb{Z}^E_2$ acts as an Enriques involution on
one $K3$ and as a generic holomorphic (not necessarily fixed-point-free)
involution on the other $K3$.
As we have shown, we cannot use the standard Weierstrass model description for $K3$ as
long as we are not in F-theory limit. Our results show that
this does not represent a problem: as long as we make sure that $\omega\rightarrow -\omega$ under the Enriques involution, 
$j$ will become symmetric in the F-theory limit. Furthermore, in this limit the standard Weierstrass model description is
equivalent to other descriptions which stay symmetric also for finite fibre volume.

\subsection*{Acknowledgments}
%\vskip -1mm
We would like to thank Bobby Acharya, Francesco Benini, Mboyo Esole, Barbara Fantechi, Stefan Groot Nibbelink, Christoph L\"udeling and Michele Trapletti for useful discussions.

\vskip 2cm

\begin{appendix}
\section{Explicit expressions for the \boldmath$\sigma_{ij}^k$}\label{appa}
{\small \begin{align}
\sigma_{14}^1  &=\frac{1}{2}\left(\pi_{14}-C_3-C_8-C_{12}-C_{15}\right)=e_3+E_4+E_8\nonumber \\
\sigma_{14}^2  &=\frac{1}{2}\left(\pi_{14}-C_4-C_7-C_{11}-C_{16}\right)=e_2+E_{12}+E_{16}\nonumber \\
\sigma_{14}^3  &=\frac{1}{2}\left(\pi_{14}-C_2-C_5-C_{9}-C_{14}\right) =e_1+e_3+e^2+e^3 \nonumber \\
&+\frac{1}{2}\left(E_1-E_2-E_3+E_4+E_5-E_6-E_7+E_8\right)\nonumber \\
&+\frac{1}{2}\left(-E_9-E_{10}-E_{11}+E_{12}-E_{13}-E_{14}-E_{15}+E_{16}\right) \nonumber \\
\sigma_{14}^4  &=\frac{1}{2}\left(\pi_{14}-C_1-C_6-C_{10}-C_{13}\right) =-e_1+e_2+e^3+e^2 \nonumber \\
&+\frac{1}{2}\left(-E_1-E_2-E_3+E_4-E_5-E_6-E_7+E_8\right)\nonumber \\
&+\frac{1}{2}\left(E_9-E_{10}-E_{11}+E_{12}+E_{13}-E_{14}-E_{15}+E_{16}\right).\\
\nonumber \\
\sigma_{23}^1 & =\frac{1}{2}\left(\pi_{23}-C_5-C_6-C_{7}-C_{8}\right)=E_7-E_6\nonumber \\
\sigma_{23}^2 & =\frac{1}{2}\left(\pi_{23}-C_1-C_2-C_{3}-C_{4}\right)=E_3-E_2\nonumber \\
\sigma_{23}^3 & =\frac{1}{2}\left(\pi_{23}-C_9-C_{10}-C_{11}-C_{12}\right)=E_{11}-E_{10}\nonumber \\
\sigma_{23}^4 & =\frac{1}{2}\left(\pi_{23}-C_{13}-C_{14}-C_{15}-C_{16}\right)=E_{15}-E_{14}\\
\nonumber \\
\sigma_{12}^1 & =\frac{1}{2}\left(\pi_{12}-C_1-C_3-C_{6}-C_{8}\right) \nonumber \\
&=\frac{1}{2}\left( -E_1-E_2+E_3+E_4-E_5-E_6+E_7+E_8\right)\nonumber \\
\sigma_{12}^2 & =\frac{1}{2}\left(\pi_{12}-C_2-C_4-C_{5}-C_{7}\right) \nonumber\\
&=e_1-e^2-e^3+\frac{1}{2}\left( E_1-E_2+E_3-E_4+E_5-E_6+E_7-E_8 \right)\nonumber \\
\sigma_{12}^3 & =\frac{1}{2}\left(\pi_{12}-C_9-C_{11}-C_{14}-C_{16}\right)\nonumber\\
&=e_1+\frac{1}{2}\left( -E_9-E_{10}+E_{11}+E_{12}-E_{13}-E_{14}+E_{15}+E_{16}\right)\nonumber \\
\sigma_{12}^4 & =\frac{1}{2}\left(\pi_{12}-C_{10}-C_{12}-C_{13}-C_{15}\right)\nonumber \\
&=-e^2-e^3+\frac{1}{2}\left( E_9-E_{10}+E_{11}-E_{12}+E_{13}-E_{14}+E_{15}-E_{16}\right)
\end{align}
\begin{align}
\sigma_{34}^1 & =\frac{1}{2}\left(\pi_{34}-C_5-C_6-C_{13}-C_{14}\right)=e_1+e^1+E_1-E_6-E_9-E_{14}\nonumber \\
\sigma_{34}^2 & =\frac{1}{2}\left(\pi_{34}-C_1-C_2-C_{9}-C_{10}\right)=e_1+e^1+E_1-E_2-E_{9}-E_{10}\nonumber \\
\sigma_{34}^3 & =\frac{1}{2}\left(\pi_{34}-C_3-C_{4}-C_{11}-C_{12}\right)=e_1+e^1-e^2-e^3+E_1+E_3-E_{9}+E_{11}\nonumber \\
\sigma_{34}^4 & =\frac{1}{2}\left(\pi_{34}-C_{7}-C_{8}-C_{15}-C_{16}\right)=e_1+e^1-e^2-e^3+E_1+E_7-E_{9}+E_{15}\\
\nonumber \\
\sigma_{13}^1 & =\frac{1}{2}\left(\pi_{13}-C_1-C_2-C_{5}-C_{6}\right)=e_2-e_3-E_2-E_6\nonumber \\
\sigma_{13}^2 & =\frac{1}{2}\left(\pi_{13}-C_9-C_{10}-C_{13}-C_{14}\right)=-E_{10}-E_{14}\nonumber \\
\sigma_{13}^3 & =\frac{1}{2}\left(\pi_{13}-C_{11}-C_{12}-C_{15}-C_{16}\right)=-e^2-e^3+e_2-e_3+E_{11}+E_{15}\nonumber \\
\sigma_{13}^4 & =\frac{1}{2}\left(\pi_{13}-C_{3}-C_{4}-C_{7}-C_{8}\right)=-e^2-e^3+E_3+E_7\\
\nonumber \\
\sigma_{42}^1 & =\frac{1}{2}\left(\pi_{42}-C_6-C_8-C_{13}-C_{15}\right)=e_2-e_3-e^3-E_5-E_6+E_{13}+E_{15}\nonumber \\
\sigma_{42}^2 & =\frac{1}{2}\left(\pi_{42}-C_5-C_7-C_{14}-C_{16}\right)=-e^3+e_2-e_3-E_6-E_8+E_{15}+E_{16}\nonumber \\
\sigma_{42}^3 & =\frac{1}{2}\left(\pi_{42}-C_2-C_4-C_{9}-C_{11}\right) = e_1+e_2-e_3-e^3 \nonumber \\
&+\frac{1}{2}\left(E_1-E_2+E_3-E_4-E_5-E_6-E_7-E_8\right)\nonumber \\
&+\frac{1}{2}\left(-E_9-E_{10}+E_{11}+E_{12}+E_{13}+E_{14}+E_{15}+E_{16}\right) \nonumber \\
\sigma_{42}^4 & =\frac{1}{2}\left(\pi_{42}-C_1-C_3-C_{10}-C_{12}\right) =-e_1+e_2-e_3-e^3 \nonumber \\
&+\frac{1}{2}\left(-E_1-E_2+E_3+E_4-E_5-E_6-E_7-E_8\right)\nonumber \\
&+\frac{1}{2}\left(E_9-E_{10}+E_{11}-E_{12}+E_{13}+E_{14}+E_{15}+E_{16}\right).
\end{align}}

\end{appendix}

\vskip 3cm
%\newpage

\end{document}